\documentclass[
 reprint,
 showkeys,
 amsmath,
 amssymb,
 aps,
 prl,
 floatfix,
 longbibliography
 ]{revtex4-2}

\usepackage{graphicx} % Required for inserting images
\usepackage[margin=2.25cm]{geometry}
\usepackage{amsmath}
\usepackage{amssymb}
\usepackage{physics}
\usepackage[dvipsnames]{xcolor}
\usepackage{hyperref}

\DeclareMathOperator{\Det}{Det}
\DeclareMathOperator{\sgn}{sgn}
\DeclareMathOperator{\arctanh}{arctanh}
\DeclareMathOperator{\diag}{diag}

\newcommand{\bgamma}[0]{\boldsymbol{\gamma}}
\newcommand{\Ay}[0]{\mathcal{A}}
\newcommand{\bAy}[0]{\boldsymbol{\mathcal{A}}}
\newcommand{\bk}{\vb{k}}

\newcommand{\sumprime}[0]{\sideset{}{'} \sum}
\newcommand{\Deltamf}[0]{\Delta_{\text{mf}}} %Change this to change all the mean-field gaps at once.

\newcommand{\citesuppmat}[0]{\cite{supp}}

\usepackage{orcidlink}

\begin{document}

\author{Benjamin A.~Levitan\,\orcidlink{0000-0003-3989-2686}\,}
\email{benjamin.levitan@usherbrooke.ca}
\author{\'Etienne Lantagne-Hurtubise\,\orcidlink{0000-0003-0417-6452}}
\affiliation{D\'epartement de physique and Institut quantique, Universit\'e de Sherbrooke, Sherbrooke, Qu\'ebec J1K 2R1, Canada}

\title{Trigonal warping enables linear optical spectroscopy\\in single-valley superconductors}

\begin{abstract}
In superconductors with multiple pairing channels, Bardasis-Schrieffer modes and clapping modes arise as fluctuations in channels whose angular momenta differ from that of the pair condensate. Crystal symmetries often impose selection rules which keep these modes optically dark. We show that if pairing occurs around a single Fermi surface, trigonal warping renders both of these modes, as well as the quasiparticle excitation gap, visible in the longitudinal and Hall optical responses. Our results suggest that rhombohedral graphene multilayers, which are believed to host the required ingredients, might offer an ideal setting for the study of exotic superconducting collective modes.
\end{abstract}

\maketitle

When multiple pairing channels compete for the title of superconducting ground state, collective order-parameter modes generically emerge. For example, in an $s$-wave superconductor, subdominant attraction in a $d$-wave channel can bind an in-gap mode corresponding to $d$-wave pair fluctuations, known as a Bardasis-Schrieffer (BaSch) mode~\cite{bardasisschrieffer1961}. Similarly, when a superconductor breaks time-reversal symmetry ($\mathcal{T}$), fluctuations in the time-reversed partner to the condensed channel manifest as generalizations of the clapping mode found in superfluid $^3$He~\cite{wolfle1976collective}. By signaling the presence of uncondensed pairing channels, collective modes carry valuable information about their host superconductor~\cite{hirashima1988collective,hirschfeld1989absorptionanisotropic,hirschfeld1992collectiveabsorption,yip1992circular,tewordt1999tripletorderparametermodes,tewordt1999collective,kee2000collective,higashitani2000order,higashitani2000response,balatsky2000collective,poniatowski2022spectroscopic,levitan2024spectroscopy,neri2025collectivetrsb,scalapino2009basch,kretzschmar2013ramanbasch,bohm2014balancing,maiti2015competing,maiti2016ramanBaSh,maiti2017raman,wu2017ramanmode,bohm2018microscopic,lee2023linear,beneklins2024selection,ponomarev2004evidence,blumberg2007MgB2leggett,hlobil2017tunneling,lee2023tunneling,nagashima2024lifshitz, levitan2024spectroscopy, nagashima2025active,niederhoff2025currentenabledopticalconductivitycollective,ponomarev2004evidence,chubukov2009ramanpnictides,huang2018identifying,sarkar2024TRSBraman, Matsushita2022}. However, BaSch and clapping modes are often invisible to linear optical response in the long-wavelength limit $\vb{q}\rightarrow0$ due to crystal symmetries \footnote{Refs.~\cite{hirschfeld1989absorptionanisotropic,hirschfeld1992collectiveabsorption,yip1992circular,higashitani2000response} argued that linear optical response could detect clapping modes because the decay of an incident electromagnetic field at the boundary of a superconductor produces $\vb{q}$ components up to the inverse penetration depth. Bulk probes at $\vb{q}=0$ remain desirable.}, motivating techniques such as Raman~\cite{scalapino2009basch,kretzschmar2013ramanbasch,bohm2014balancing,maiti2015competing,maiti2016ramanBaSh,maiti2017raman,wu2017ramanmode,bohm2018microscopic,lee2023linear,beneklins2024selection} or tunneling spectroscopy~\cite{ponomarev2004evidence,hlobil2017tunneling,lee2023tunneling}.

Previous studies have shown how linear electromagnetic coupling to superconducting collective modes can survive the long-wavelength limit in certain multi-band superconductors~\cite{kamatani2022optical,nagashima2024lifshitz,levitan2024spectroscopy, nagashima2025active,niederhoff2025currentenabledopticalconductivitycollective, matsushita2026microwavekerrfaradayresonancetwodimensional}, for example, or when the gap function breaks in-plane rotational symmetries~\cite{poniatowski2022spectroscopic}. An alternative is to apply a supercurrent \cite{moor2017movinghiggs,nakamura2019supercurrentactivation, crowley2022supercurrent,wang2025higgsdisorder}: Cooper pairs then acquire a net velocity, coupling light to order-parameter amplitude fluctuations (i.e., the Higgs mode) even at $\vb{q}=0$.

\begin{figure}[t]
\centering
\includegraphics[width=\columnwidth]{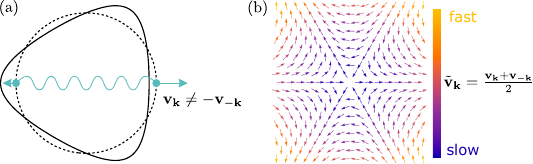}
\caption{\label{fig:warping} Trigonal warping of a single Fermi surface (a) corresponds to an inversion-even contribution $\tilde{\vb{v}}$ to the electron velocity (b). Zero-momentum Cooper pairs thus acquire a non-zero velocity, enabling linear optical response in the superconducting state.}
\end{figure}

In this Letter, we show how trigonal warping enables a similar effect for BaSch and clapping modes when pairing occurs around a single Fermi surface (see Fig.~\ref{fig:warping}). These modes then become visible to any probe sensitive to the frequency-dependent superfluid stiffness, including longitudinal and Hall conductivity, polar Kerr effect~\cite{Kapitulnik2009polar}, microwave reflectometry and related circuit-QED techniques~\cite{Boettcher2024, Kreidel2024inductance, Banerjee2025superfluid, Tanaka2025superfluid, Jha2025tunable, zaman2025kineticinductancefewlayernbse2, kreidel2025observingunconventionalsuperconductivitykinetic}. Crucially, the optical couplings enabled by trigonal warping do not require any background field or supercurrent.

Rhombohedral four-, five-, and six-layer graphene (R$n$G) may provide precisely the ingredients required for the optical couplings we discuss. Recent experiments~\cite{Han2025, morissette2025superconductivity, nguyen2025hierarchytopologicalsuperconductingstates} have revealed several superconducting phases that emerge from spin-and-valley-polarized normal states, each hosting a single $C_3$-symmetric Fermi surface. Theory suggests that some of these superconductors are chiral~\cite{ wang2024chiralsuperconductivityparentchern, geier2024chiraltopologicalsuperconductivityisospin, qin2024chiralfinitemomentumsuperconductivitytetralayer, yang2024topologicalincommensurate, maymann2025pairing, chou2025intravalley, jahin2025enhancedkohnluttingertopologicalsuperconductivity, yoon2025quartermetalsuperconductivity, parramartinez2025bandrenormalizationquartermetals, christos2025finitemomentumpairingsuperlatticesuperconductivity, kim2025topologicalchiral, shavit2025quantum, kim2025variational, patri2025familymultilayergraphenesuperconductors, chen2025intrinsicsuperconductingdiodeeffect, Murshed2025, Gaggioli2025, chou2025superconductivityphononmediatedretardationsingleflavor, karuzin2026chiraltwobodyboundstates}. They might therefore host clapping modes (in addition to BaSch modes). We discuss the implications of our results for R$n$G, and propose microwave impedance spectroscopy as an ideal probe of its superconducting collective modes.

\emph{Model}---We consider a model of spinless electrons with pairing interactions,
\begin{equation}
    H
    =
    \sum_{\vb{k}} \xi_{\vb{k}} c^{\dag}_{\vb{k}} c_{\vb{k}}
    + \frac{1}{2 L^2} \sumprime_{\vb{k k' q}} V_{\vb{k k'}}
    c^{\dag}_{\vb{k+q}} 
    c^{\dag}_{-\vb{k}}
    c_{-\vb{k'}}
    c_{\vb{k'+q}},
\end{equation}
where $c_{\vb{k}}$ annihilates an electron with momentum $\vb{k}$ relative to an arbitrary origin (e.g., the valley momentum in graphene), and $L$ is the linear system dimension. The primed sum indicates that electrons participating in the interaction live within a cutoff $\Lambda$ of the Fermi level, as usual in BCS-like theories. We remain agnostic to the microscopic origin of the interactions, and therefore to the physical interpretation of $\Lambda$. It will be useful to work with the even and odd parts of the dispersion relation,
\begin{equation}
    \xi_{\vb{k}}
    =
    \underbrace{
        \frac{\xi_{\vb{k}} + \xi_{-\vb{k}}}{2}
    }_{\xi_{0,\vb{k}}}
    +
    \underbrace{
        \frac{\xi_{\vb{k}} - \xi_{-\vb{k}}}{2}
    }_{\tilde{\xi}_{\vb{k}}},
\end{equation}
and the corresponding odd and even parts of the electron velocity
\begin{equation}
    \vb{v}_{\vb{k}} = \partial_{\vb{k}} \xi_{\vb{k}}
    =
    \underbrace{
        \frac{\vb{v}_{\vb{k}}-\vb{v}_{-\vb{k}}}{2}
    }_{\vb{v}_{0,\vb{k}}}
    +
    \underbrace{
        \frac{\vb{v}_{\vb{k}}+\vb{v}_{-\vb{k}}}{2}
    }_{\tilde{\vb{v}}_{\vb{k}}}.
\end{equation}

We assume that the interactions comprise two pairing channels $\ell = 1, 2$:
\begin{equation}    \label{eq:interaction}
    V_{\vb{k k'}}
    =
    - \frac{1}{2} \sum_{\ell = 1, 2} g_{\ell} \chi^{(\ell)}_{\vb{k}} \overline{\chi}^{(\ell)}_{\vb{k'}},
\end{equation}
with attractive couplings $g_1 \ge g_2 > 0$; the factor of $1/2$ is for convenience. (Including more channels is conceptually straightforward.) $\delta g = g_1 - g_2$ quantifies the competition between pairing channels at the critical temperature; $\delta g = 0$ means perfect competition.

Passing over to a functional-integral approach, we decompose interactions via a standard Cooper-channel Hubbard-Stratonovich transformation~\cite{altlandandsimons2023}. Before coupling to electromagnetism, the action is
\begin{subequations}    \label{eq:exact_HS_action}
\begin{equation}
    S
    =
    S_{\Psi}
    +
    S_{\Psi \Delta}
    +
    \frac{L^2 \beta}{2} \sum_{q \ell}
    \frac{|\Delta^{(\ell)}_q|^2}{g_{\ell}},
\end{equation}
with
\begin{equation}
    S_{\Psi}
    =
    -\frac{\beta}{2} \sum_k \Psi^{\dag}_k
    \begin{pmatrix}
        i \omega_n - \xi_{\vb{k}} & 0 \\
        0 & i \omega_n + \xi_{-\vb{k}}
    \end{pmatrix}
    \Psi_k
\end{equation}
and
\begin{equation}
    S_{\Psi \Delta}
    =
    \\
    - \frac{\beta}{2} \sumprime_{k q \ell}
    \Psi^{\dag}_{k+q}
    \begin{pmatrix}
        0 & \Delta^{(\ell)}_{q} \chi^{(\ell)}_{\vb{k}} \\
        \overline{\Delta}^{(\ell)}_{-q} \overline{\chi}^{(\ell)}_{\vb{k}} & 0
    \end{pmatrix}
    \Psi_k.
\end{equation}
\end{subequations}
Here we introduced the Nambu representation $\Psi^{\dag}_k = \begin{pmatrix} \overline{c}_k & c_{-k} \end{pmatrix}$, which double-counts degrees of freedom in our model. $k = (i \omega_n, \vb{k})$ and $q = (i \Omega_m, \vb{q})$ respectively combine fermionic and bosonic Matsubara frequencies with spatial momenta. $\Delta^{(\ell)}_q$ is the fluctuating pairing field in channel $\ell$; an electron with momentum $\vb{k}$ feels a fluctuating gap $\Delta_{\vb{k}, q} = \sum_{\ell} \Delta^{(\ell)}_q \chi^{(\ell)}_{\vb{k}}$. 

The gap equation arises as the saddle-point condition for the $\Delta^{(\ell)}$ fields after integrating out the fermions. Our analysis holds when inversion breaking is weak enough that $|\tilde{\xi}_{\vb{k}}| < E_{0, \vb{k}} \equiv \sqrt{\xi_{0,\vb{k}}^2 + |\Delta_{\vb{k}}|^2}$ at all relevant momenta, where $\Delta_{\vb{k}}$ is the mean-field gap function (i.e., the saddle-point value of $\Delta_{\vb{k},q=0}$). For a nodeless state with strong pairing interactions, this condition can be satisfied self-consistently. In that case, surprisingly, $\tilde{\xi}$ has no impact on pairing and disappears from all calculations in the long-wavelength limit at $T \rightarrow0$~\citesuppmat, except for its contribution $\propto \tilde{\vb{v}}_{\vb{k}} \tau^z$ to the paramagnetic vertex  (more on this below). 

We assume that the $\ell = 1$ channel condenses, producing the pairing gap
\begin{equation}
    \Delta_{\vb{k}} 
    = \Deltamf \chi^{(1)}_{\vb{k}}.
\end{equation}
We take $\Deltamf > 0$ without loss of generality.

\emph{Collective modes and electromagnetic response}---Next we parametrize the pairing fields in terms of their fluctuations around the mean-field configuration. In spacetime coordinates $x = (i \tau, \vb{r})$, we write
\begin{subequations}    \label{eq:fluctuation_coordinates}
\begin{align}
    \Delta^{(1)}(x)
    =&
    e^{i \theta(x)} \left[
        \Deltamf + h(x)
    \right] 
    , \\
    \Delta^{(2)}(x)
    =&
    e^{i \theta(x)} \left[
        a(x) + i b(x)
    \right].
\end{align}
\end{subequations}
$h(x)$ and $\theta(x) \in \mathbb{R}$ are respectively the Higgs and  Anderson-Bogoliubov-Goldstone modes. The interpretation of $a(x)$ and $b(x) \in \mathbb{R}$ depends on the pairing channels chosen. If $\chi^{(1)}_{\vb{k}} = \overline{\chi}^{(2)}_{\vb{k}}$, i.e., if the two pairing channels are time-reversed partners, then $a$ and $b$ are clapping modes. If instead $\chi^{(1)}_{\vb{k}}$ and $\chi^{(2)}_{\vb{k}}$ come from unrelated irreps, e.g., $p+ip$ and $f+if$ (or $s\text{-wave}$ and $d\text{-wave}$ in a spin-singlet model), then $a$ and $b$ would instead constitute BaSch modes. It will be useful to collect the fields $h, a$ and $b$ into $\Phi = \begin{pmatrix} h & a & b \end{pmatrix}^T$.

We now couple to electromagnetism through the standard minimal substitution. Electromagnetic linear response is controlled by the effective action at bilinear order in the gauge potential $A = (A_0, \vb{A})$, so we drop all terms at third order or higher. We may remove the overall order-parameter phase $\theta(x)$ by a gauge transformation~\cite{sharapov2001effectivetheory,lutchyn2008gaugeinvariantchiral,boyack2020multiplemodes}, which combines $A$ and the spacetime-gradient of $\theta$ into the gauge-invariant field $\Ay = (\Ay_0, \bAy) \equiv (A_0 + \frac{1}{2e} \partial_{\tau} \theta, \vb{A} - \frac{1}{2e}  \grad \theta)$. Note that $\theta$ will ultimately play no role in our calculation when we focus on $\vb{q}=0$ optical response~\citesuppmat. We then reorganize the action for an expansion around mean field,
\begin{multline}
    S
    =
    L^2 \beta \sum_q \Phi^T_{-q} \hat{g}^{-1} \Phi_q
    \\
    - \frac{\beta}{2} \sum_{k q}
    \Psi^{\dag}_{k+q} \left(
        \delta_{q,0} \mathcal{G}^{-1}_k
        - \mathbb{V}_{k+q, k}
    \right) \Psi_k ,
\end{multline}
where $\hat{g} = \diag(g_1, g_2, g_2)$,
\begin{equation}
    \mathcal{G}^{-1}_k
    =
    \begin{pmatrix}
        i \omega_n - \xi_{\vb{k}} & \Delta_{\vb{k}} \\
        \overline{\Delta}_{\vb{k}} & i \omega_n + \xi_{-\vb{k}}
    \end{pmatrix}
\end{equation}
is the mean-field inverse fermion propagator, and
\begin{equation}
    \mathbb{V}
    =
    \mathbb{V}^{(\Ay_0)}
    + \mathbb{V}^{(\bAy)}
    + \mathbb{V}^{(\bAy^2)}
    + \mathbb{V}^{(\Phi)} 
\end{equation}
collects the vertices coupling the fermions to each of the other fields. The collective-mode contribution to the superfluid stiffness results from the order-parameter fluctuation vertex
\begin{subequations} \label{eq:vertices}
\begin{multline}
    \mathbb{V}^{(\Phi)}_{k+q, k} =
    \\
    -\begin{pmatrix}
        0 
        & \chi^{(1)}_{\vb{k}} h_q 
        + \chi^{(2)}_{\vb{k}} (a_q + i b_q) \\
        \overline{\chi}^{(1)}_{\vb{k}} h_q
        + \overline{\chi}^{(2)}_{\vb{k}} (a_q - i b_q) 
        & 0
    \end{pmatrix},
\end{multline}
and the paramagnetic vertex
\begin{equation}
    \mathbb{V}^{(\bAy)}_{k+q, k}
    =
    - e \bAy_q \vdot \begin{pmatrix}
        \bgamma_{\vb{k+q},\vb{k}} & 0 \\
        0 & -\bgamma_{\vb{-k},\vb{-k-q}}
    \end{pmatrix}.
\end{equation}
\end{subequations}
At $\vb{q}=0$, the paramagnetic vertex function is simply the velocity, $\vb{\gamma}_{\vb{k},\vb{k}} = \vb{v}_{\vb{k}}$. The expression at $\vb{q}\ne0$ is cumbersome for all but the simplest dispersion relations $\xi_{\vb{k}}$; see the Supplemental Material~\citesuppmat, where we also provide the scalar-potential and diamagnetic vertices $\mathbb{V}^{(\Ay_0)}$ and $\mathbb{V}^{(\bAy^2)}$.

Integrating out the fermion and expanding to second order in $\mathbb{V}$ yields the effective action
\begin{multline}    \label{eq:gpf_action}
    S_{\text{eff}}
    = 
    L^2 \beta \sum_q \Big[
        -\Phi^T_{-q}
        \hat{\mathcal{D}}^{-1}_q
        \Phi_q
        \\
        + K^{\mu \nu}_{\text{f},q} \Ay_{\mu,-q} \Ay_{\nu,q}
        + \Ay_{\mu, -q} (\Pi^{\mu}_q)^T \Phi_q
    \Big],
\end{multline}
with implicit summation on $\mu, \nu \in \lbrace 0, x, y \rbrace$. $\hat{\mathcal{D}}^{-1}_q$ is the (matrix) inverse propagator for order-parameter fluctuations. $K_{\text{f}}^{00}$ and $K_{\text{f}}^{ij}$ are the fermionic contribution to the compressibility and the superfluid stiffness tensor, respectively. We provide explicit expressions in~\citesuppmat. $\Pi^{\mu} = \begin{pmatrix} \Pi^{\mu h} & \Pi^{\mu a} & \Pi^{\mu b} \end{pmatrix}^T$ collects the bilinear couplings between $\mathcal{A}_{\mu}$ and the order-parameter modes. 

Integrating out the collective modes renormalizes the electromagnetic response tensor:
\begin{equation}    \label{eq:response_renormalization}
    K^{\mu \nu}_q 
    = K^{\mu \nu}_{\text{f}, q}
    + 
        \frac{1}{4} (\Pi^{\mu}_{-q})^T \hat{\mathcal{D}}_q \Pi^{\nu}_q
    ,
\end{equation}
which can be analytically continued to real frequency, $i\Omega_m \rightarrow \Omega + i 0^+$. When the coupling $\Pi^{j \phi}_q$ is non-vanishing in the long-wavelength limit $\vb{q} \rightarrow 0$, the corresponding mode $\phi$ appears in the renormalized superfluid stiffness and therefore the optical response. 

We now evaluate the couplings in the $\vb{q} \rightarrow 0$ limit. The current-density contribution from the fluctuations $\Phi$ (i.e., the quantity coupling to $\Ay_{j,-q}$ in Eq.~\eqref{eq:gpf_action}) is
\begin{multline}    \label{eq:mode_couplings_diagram}
    -(\Pi^j_q)^T \Phi_q
    \xrightarrow{\vb{q}
    \rightarrow0}
    \\
    \frac{e}{L^2 \beta} \sumprime_k
    \tr{
        \mathcal{G}_{k}
        (v^j_{0, \vb{k}} \tau^0 + \tilde{v}^j_{\vb{k}} \tau^z)
        \mathcal{G}_{k+q}
        \mathbb{V}^{(\Phi)}_{k+q, k}
    }.
\end{multline}
Here $\tau^0$ and $\tau^z$ are the identity and third Pauli matrix in Nambu space. Equation~\eqref{eq:mode_couplings_diagram} reveals an analogy to supercurrent-enabled optical coupling~\cite{moor2017movinghiggs,nakamura2019supercurrentactivation,allocca2019cavitysuperconductorpolaritons, niederhoff2025currentenabledopticalconductivitycollective,wang2025higgsdisorder}: $\tilde{\vb{v}}_{\vb{k}}$ acts similarly to a background supercurrent at velocity $\vb{v}_s$, since both are inversion-even contributions to the electronic velocity which therefore contribute a $\tau^z$ to the paramagnetic vertex. Performing the Matsubara integral at $T \rightarrow 0$, we find
\begin{multline}    \label{eq:mode_couplings_general}
    \Pi^{j}_{\vb{q}=0}(i\Omega_m)
    =
    -e \int' \frac{\dd^2\vb{k}}{(2\pi)^2}
    \frac{2 \Deltamf}{E_{0,\vb{k}}}
    \frac{\tilde{v}^j_{\vb{k}}}{(i \Omega_m)^2 - 4 E_{0,\vb{k}}^2}
    \\
    \begin{pmatrix}
        2 \xi_{0,\vb{k}} \left|\chi^{(1)}_{\vb{k}}\right|^2
        \\
        \Re \left[
            \overline{\chi}^{(1)}_{\vb{k}}
            \chi^{(2)}_{\vb{k}}
            ( 2 \xi_{0,\vb{k}} + i \Omega_m )
        \right]
        \\
        - \Im \left[
            \overline{\chi}^{(1)}_{\vb{k}}
            \chi^{(2)}_{\vb{k}}
            ( 2 \xi_{0,\vb{k}} + i \Omega_m )
        \right]
    \end{pmatrix},
\end{multline}
where the 3-tuple is indexed by $\phi = h, a, b$. Equation~\eqref{eq:mode_couplings_general} assumes only that $|\tilde{\xi}| < E_{0,\vb{k}} = \sqrt{\xi_{0,\vb{k}}^2 + |\Delta_{\vb{k}}|^2}$ everywhere, and that $\Delta_{\vb{k}} = \Deltamf \chi^{(1)}_{\vb{k}}$. 

Unlike a constant background velocity, $\tilde{\vb{v}}_{\bk}$ (which dictates the Cooper pair velocity) depends on direction. This is the key to enabling optical couplings to the $a$ and $b$ modes: the angle dependence of $\tilde{\vb{v}}_{\bk}$ can compensate that of $\overline{\chi}^{(1)}_{\bk} \chi^{(2)}_{\bk}$, which 
a supercurrent cannot do. Conversely, while supercurrent optically activates the Higgs mode~\cite{moor2017movinghiggs,nakamura2019supercurrentactivation, crowley2022supercurrent,wang2025higgsdisorder}, warping of the Fermi surface cannot since $\tilde{\vb{v}}_{\vb{k}}$ has a vanishing angular average (assuming that $\xi_{\bk}$ respects any rotational symmetry $C_{n}$).

\emph{Toy model---}For concreteness, we now assume an otherwise-isotropic dispersion relation with trigonal warping, $\tilde{\xi}_{\vb{k}} = \eta \left( \frac{1}{3} k_x^3 - k_x k_y^2 \right)$. We choose $p \pm i p$ pairing channels with Fermi-surface-projected form factors 
\begin{equation}
    \chi^{(1)}_{\vb{k}}
    =
    \overline{\chi}^{(2)}_{\vb{k}}
    =
    e^{i \varphi_{\vb{k}}},
\end{equation}
where $\varphi_{\vb{k}}$ denotes the angle of $\vb{k}$ relative to the $k_x > 0$ axis. With this choice, $\delta g$ quantifies $\mathcal{T}$ breaking in the normal state. At $\delta g = 0$ the model spontaneously breaks $\mathcal{T}$, and at $\delta g > 0$ the $\ell=1$ ($p + ip$) channel is selected explicitly. The inversion-even $\tilde{\vb{v}}$ comes purely from trigonal warping,
\begin{equation}
    \tilde{\vb{v}}_{\vb{k}} 
    = \partial_{\vb{k}} \tilde{\xi}_{\vb{k}}
    = \eta \vb{k}^2 (\cos 2 \varphi_{\vb{k}}, -\sin 2 \varphi_{\vb{k}}).
\end{equation}

Consider the couplings in Eq.~\eqref{eq:mode_couplings_general}. As explained above, the Higgs component vanishes since $\tilde{\vb{v}}_{\vb{k}}$ averages to zero over the Fermi surface, and the rest of the integrand is rotationally invariant. However, the clapping mode components include the factor $\overline{\chi}^{(1)}_{\vb{k}} \chi^{(2)}_{\vb{k}} = e^{-2 i \varphi_{\vb{k}}}$ which balances the angular dependence of $\tilde{\vb{v}}_{\vb{k}}$. Those modes become optically active:
\begin{subequations}    \label{eq:couplings}
\begin{multline}
    \Pi^{x a}_{\vb{q}=0}(i\Omega_m)
    =
    -\Pi^{y b}_{\vb{q}=0}(i\Omega_m)
    =
    \\
    2 e \nu \eta m \Deltamf \left(
        \log \frac{2 \Lambda}{\Deltamf}
        - z \cot z
    \right)
\end{multline}
and
\begin{equation}
    \Pi^{x b}_{\vb{q}=0}(i\Omega_m)
    =
    \Pi^{y a}_{\vb{q}=0}(i\Omega_m)
    =
     2 i e \nu \eta  m \mu z \sec z , 
\end{equation}
\end{subequations}
where $\nu$ is the Fermi-level density of states corresponding to $\xi_{0,\vb{k}}$, and
\begin{equation}
    z = \arcsin \frac{i\Omega_m}{2\Delta}.
\end{equation}
The calculation is virtually unchanged if we instead choose $p+ip$ and $f+if$ form factors, $\chi^{(1)} = e^{i \varphi}$ and $\chi^{(2)} = e^{3 i \varphi}$, since then $\overline{\chi}^{(1)} \chi^{(2)} = e^{-2 i \varphi}$. The same reasoning holds for any pair of form factors differing by two units of angular momentum, $|L - L_{\text{gs}}| = 2$. In our simple model, modes with $|L - L_{\text{gs}}| > 2$ remain dark since $\xi_0$ has complete rotational symmetry and $\tilde{\vb{v}}$ has twofold rotational symmetry. However, on symmetry grounds, one expects from Eq.~\eqref{eq:mode_couplings_general} that if $\xi_0$ contains hexagonal warping, then modes with $|L - L_{\text{gs}}| = 4$ will become bright. Similar reasoning applies for further angular harmonics consistent with $C_3$ symmetry.

The real-frequency bosonic propagator $\hat{\mathcal{D}}_{\vb{q}=0}(\Omega+i0^+)$ has a pole at frequency $\Omega_c$, which appears in the superfluid stiffness as per Eq.~\eqref{eq:response_renormalization}. In the limit where $\delta g \ll g_1$, we find
\begin{equation}
    \Omega_c
    \approx
    \sqrt{2} \Deltamf \left(
        1 + \frac{2}{\pi}
        \frac{1}{\nu g_1} \frac{\delta g / g_1}{1 - \delta g / g_1}
    \right).
\end{equation}
We provide an explicit expression for the bosonic propagator in~\citesuppmat, along with the quasiparticle contribution to the superfluid stiffness.

\begin{figure}[t]
    \centering
    \includegraphics[width=\columnwidth]{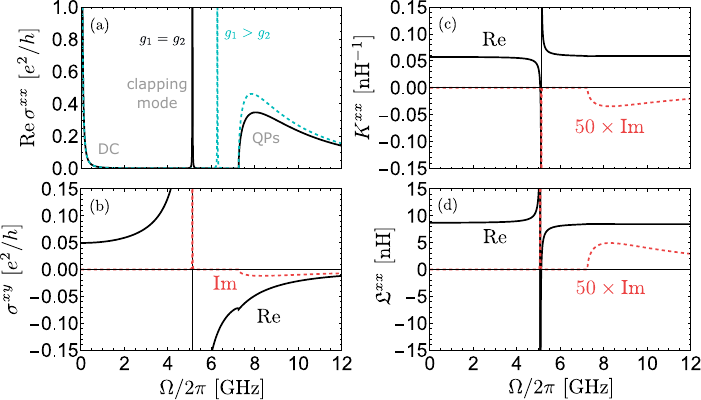}
    \caption{\label{fig:response} Linear optical response of a single-valley $p+ip$ superconductor with trigonal warping. (a) Real part of longitudinal conductivity. (b) Hall conductivity. (c) Longitudinal superfluid stiffness. (d) Longitudinal inductance of a square sample. $g_{\ell}$ is the pairing strength in channel $\ell=1,2$. All panels use parameters inspired by rhombohedral graphene~\citesuppmat: $\mu = 1$ meV, $\nu = 5$ eV$^{-1}$nm$^{-2}$, $\eta = 20$ meV nm$^3$, $\Deltamf = 15$ $\mu$eV, and $\Lambda = 150$ $\mu$eV. We give the frequency $\Omega / 2 \pi$ a small imaginary part, $1.5 \times 10^{-4}$ GHz.}
\end{figure}

In Fig.~\ref{fig:response}, we show the renormalized optical conductivity tensor
\begin{equation}
    \sigma^{ij}(\Omega) = \frac{i K^{ij}_{\vb{q}=0}(\Omega + i 0^+)}{\Omega + i 0^+},
\end{equation}
along with the longitudinal stiffness, $K^{xx}_{\vb{q}=0}$, and the longitudinal inductance of a square sample, $\mathfrak{L}^{xx}$ (see \citesuppmat). The real part of the longitudinal conductivity consists of $\delta$-functions at DC ($\Omega = 0$) and the clapping mode frequency $\Omega_c$, and a broad feature above the quasiparticle gap $2 \Deltamf$. Breaking the degeneracy between pairing channels ($g_1 > g_2$) pushes the clapping mode up in energy from $\sqrt{2} \Deltamf$. (Within our model, the $a$ and $b$ modes are degenerate for all values of $\delta g$, so they contribute a single pole; they may split when the gap function breaks rotational symmetry~\cite{poniatowski2022spectroscopic}). The Hall conductivity also exhibits distinctive features at $\Omega_c$ and $2 \Deltamf$. In clean superconductors, inversion symmetry typically prevents optical excitation of quasiparticles across the gap~\cite{xunonlinear2019, ahn2021theory,papaj2022supercurrent,crowley2022supercurrent} and enforces a vanishing Hall response~\cite{lutchyn2009anomalous, taylor2012intrinsic, denys2021origin}. By breaking inversion symmetry, trigonal warping activates both of these responses in our model.

\emph{Application to  R$n$G}---In rhombohedral graphene with four, five and six layers, experiments have reported superconductors that condense from spin-and-valley-polarized normal states~\cite{Han2025, morissette2025superconductivity, nguyen2025hierarchytopologicalsuperconductingstates}, suggesting that pairing occurs within a single flavor. Fermion anticommutation then forces the gap function to be either chiral or nodal. Theoretical models often prefer a chiral solution as it admits a full gap, which increases the condensation energy relative to a nodal state~\cite{patri2025familymultilayergraphenesuperconductors, maymann2025pairing, chou2025intravalley, geier2024chiraltopologicalsuperconductivityisospin, qin2024chiralfinitemomentumsuperconductivitytetralayer, yang2024topologicalincommensurate, jahin2025enhancedkohnluttingertopologicalsuperconductivity, yoon2025quartermetalsuperconductivity, parramartinez2025bandrenormalizationquartermetals, christos2025finitemomentumpairingsuperlatticesuperconductivity, kim2025topologicalchiral, shavit2025quantum, kim2025variational}. Since the Fermi surface in each valley is trigonally warped, R$n$G likely meets the criteria we have discussed for optical coupling to condensate modes. 

In the SM we provide estimates for parameters appropriate for R$n$G, used in Fig.~\ref{fig:response}. In particular, we predict a kinetic inductance on the order of tens of nH, at characteristic frequencies of a few GHz---natural scales for microwave reflectometry or circuit-QED inspired techniques. Recent measurements in few-layer NbSe$_2$~\cite{Kreidel2024inductance, zaman2025kineticinductancefewlayernbse2}, MoTe$_2$~\cite{kreidel2025observingunconventionalsuperconductivitykinetic} and moire graphene~\cite{Tanaka2025superfluid, Banerjee2025superfluid, Jha2025tunable} have reported kinetic inductances on a similar scale.

If a chiral superconductor \emph{spontaneously} breaks $\mathcal{T}$ (i.e., if it arises from a $\mathcal{T}$-symmetric normal state), one generically expects a clapping mode comprising fluctuations in the time-reversed partner to the condensed channel. In superconductors descended from $\mathcal{T}$-broken normal states (like R$n$G), the band-projected pairing interactions will also break $\mathcal{T}$, and attraction in the time-reversed channel could be much weaker than in the condensed channel. Another pairing channel (in a different irrep) could also have comparable attraction to the ground state~\cite{maymann2025pairing, karuzin2026chiraltwobodyboundstates}. As a result, a BaSch mode might arise at lower energy than the clapping mode.

Identifying the character of any collective mode observed in experiment could help gain insight into the underlying pairing interactions. The BaSch modes we considered change pair angular momentum by two units while preserving its direction, while the clapping mode reverses the pair angular momentum. Therefore, although the optical responses of these modes are identical in our model, we expect them to couple differently to an out-of-plane magnetic field. Future work will study this possibility within microscopic models. 

\emph{Conclusion}---We have demonstrated that the clapping modes of a chiral $p$-wave superconductor are optically active when the parent normal state hosts a single Fermi surface with trigonal warping. The same is true for the subset of Bardasis-Schrieffer modes which change Cooper pair angular momentum by two units. The physical origin of this effect is that trigonal warping breaks the antisymmetry in velocity between the two electrons forming a Cooper pair, activating couplings between order-parameter collective modes and light. Notably, order-parameter modes become visible as poles in the longitudinal and transverse components of the AC superfluid stiffness. 

Our results extend the zoo of collective modes detectable in linear optical spectroscopy. We argued that valley-polarized superconductors in R$n$G are a promising platform to test our ideas. Future work will consider more realistic models of R$n$G and investigate how experimental tuning parameters---such as charge density, perpendicular displacement fields, and magnetic fields---affect their collective mode spectra.

\emph{Acknowledgments---}We thank Erez Berg for fruitful discussions and collaborations on related topics. BAL gratefully acknowledges support from the Canada First Research Excellence Fund (CFREF). ELH is grateful for support from the National Sciences and Engineering Council of Canada (NSERC), grant RGPIN-2025-06136, and start-up funds from the Faculté des Sciences at Université de Sherbrooke.

\bibliography{biblio,collectivemodes}

\begin{widetext}
\section{Supplemental material}
\subsection{1. Fourier conventions}
For fermion operators, we use the finite-volume normalization of the spatial Fourier transform
\begin{equation}
    \psi(\vb{r})
    =
    \frac{1}{\sqrt{L^2}} \sum_{\vb{k}}
    e^{i \vb{k} \vdot \vb{r}} c_{\vb{k}}
    \qquad \text{and} \qquad
    c_{\vb{k}}
    =
    \frac{1}{\sqrt{L^2}} \int \dd^d \vb{r} \, 
    e^{-i \vb{k} \vdot \vb{r}} \psi(\vb{r}),
\end{equation}
corresponding to density (number) normalization in coordinate (momentum) space,
\begin{equation}
    \lbrace \psi(\vb{r}), \psi^{\dag}(\vb{r'}) \rbrace
    =
    \delta(\vb{r-r'})
    \qquad \text{and} \qquad
    \lbrace c_{\vb{k}}, c^{\dag}_{\vb{k'}} \rbrace
    =
    \delta_{\vb{k k'}}.
\end{equation}
For the imaginary-time dependence we use
\begin{equation}
    c_{k \equiv (i \omega_n, \vb{k})}
    =
    \frac{1}{\beta} \int_0^{\beta} \dd{\tau}
    e^{i \omega_n \tau} c_{\vb{k}}(\tau)
    \qquad \text{and} \qquad
    c_{\vb{k}}(\tau)
    =
    \sum_{i \omega_n} e^{-i \omega_n \tau} c_k
\end{equation}
for fermions, with Matsubara frequencies $\omega_n = (2n+1) \pi /\beta$. We use the analogous convention for bosons, only with bosonic Matsubara frequencies $\Omega_m = 2n \pi / \beta$.

\subsection{2. Concrete model}
When a concrete electronic dispersion relation is required, we take the simple form
\begin{equation}
    \xi_{\vb{k}}
    = \frac{1}{2m} |\vb{k}|^2 
    + \eta \left( \frac{1}{3} k_x^3 - k_x k_y^2 \right) 
    + \alpha |\vb{k}|^4 - \mu.
    \label{eq:app_dispersion_trigonal}
\end{equation}
relative to the chemical potential $\mu$. $m$ is the electron effective mass. $\eta$ quantifies trigonal warping. We include the term $\alpha |\vb{k}|^4$ only for stability; we will not use it explicitly in any calculations. The next order of allowed terms, at $O(|\vb{k}|^6)$, would contribute hexagonal warping, breaking the full [$U(1)$] rotational symmetry of $\xi_{0,\vb{k}}$ down to $C_6$. Including those terms would complicate the angular integrals throughout this work, so we neglect them for simplicity. When concrete choices of pairing channels are required, we take $U(1)$-symmetric $p \pm ip$ form factors
\begin{equation}
    \chi^{(1)}_{\vb{k}} = \overline{\chi}^{(2)}_{\vb{k}}
    = e^{i \varphi_{\vb{k}}}.
\end{equation}
If the normal state respects time-reversal symmetry, then $g_1 = g_2$. If it does not, then $g_1 \ne g_2$ in general. We take $g_1 \ge g_2$.

\subsection{3. Hubbard-Stratonovich transformation}
Decoupling by Hubbard-Stratonovich in the Cooper channel~\cite{altlandandsimons2023} introduces one complex auxiliary boson $\Delta^{(\ell)}$ per pairing channel $\ell$.
The result is
\begin{equation}
    S[c, \Delta]
    =
    \int_{\tau} \sum_{\vb{k}} \overline{c}_{\vb{k}} 
    (\partial_{\tau} + \xi_{\vb{k}}) c_{\vb{k}}
    +
    \int_{\tau} \sum_{\vb{q} \ell} 
    \frac{L^2}{g_{\ell}} |\Delta^{(\ell)}_{\vb{q}}|^2
    -
    \frac{1}{2} \int_{\tau} \sumprime_{\vb{k q} \ell}
    \left(
        \Delta^{(\ell)}_{\vb{q}} \chi^{(\ell)}_{\vb{k}} \overline{c}_{\vb{k+q}} \overline{c}_{-\vb{k}}
        + \overline{\Delta}^{(\ell)}_{\vb{q}} \overline{\chi}^{(\ell)}_{\vb{k}} c_{-\vb{k}} c_{\vb{k+q}}
    \right)
\end{equation}
where $\int_{\tau} \equiv \int_0^{\beta} \dd{\tau}$. To arrive at the Nambu representation, we first rewrite
\begin{equation}
    \sumprime_{\vb{k q}}
    \overline{\Delta}^{(\ell)}_{\vb{q}} \overline{\chi}^{(\ell)}_{\vb{k}} c_{-\vb{k}} c_{\vb{k+q}}
    =
    - \sumprime_{\vb{k q}}
    \overline{\Delta}^{(\ell)}_{\vb{q}} \overline{\chi}^{(\ell)}_{\vb{k}} c_{\vb{k+q}} c_{-\vb{k}} 
    =
    - \sumprime_{\vb{k q}}
    \overline{\Delta}^{(\ell)}_{-\vb{q}} 
    \overline{\chi}^{(\ell)}_{-\vb{k}} 
    c_{\vb{-k-q}} c_{\vb{k}}
    =
    \sumprime_{\vb{k q}}
    \overline{\Delta}^{(\ell)}_{-\vb{q}} 
    \overline{\chi}^{(\ell)}_{\vb{k}} 
    c_{\vb{-k-q}} c_{\vb{k}}
\end{equation}
where the last equality used $\chi^{(\ell)}_{-\vb{k}} = -\chi^{(\ell)}_{\vb{k}}$ as appropriate in a model of spinless fermions. Fourier transforming with respect to imaginary time, we then obtain Eq.~\eqref{eq:exact_HS_action} of the main text,
\begin{equation}    \label{eq:HS_action}
    S[\Psi, \Delta]
    =
    \frac{\beta}{2} \sum_{k}
    \Psi^{\dag}_k \begin{pmatrix}
        -i \omega_n + \xi_{\vb{k}} & 0 \\
        0 & -i \omega_n - \xi_{-\vb{k}}
    \end{pmatrix} \Psi_k
    +
    L^2 \beta \sum_{q \ell} \frac{1}{g_{\ell}} |\Delta^{(\ell)}_{q}|^2
    -
    \frac{\beta}{2} \sumprime_{k q \ell}
    \Psi^{\dag}_{k+q} \begin{pmatrix}
        0 & \Delta^{(\ell)}_q \chi^{(\ell)}_{\vb{k}} \\
        \overline{\Delta}^{(\ell)}_{-q} \overline{\chi}^{(\ell)}_{\vb{k}} & 0
    \end{pmatrix} \Psi_{k}
\end{equation}
with $\Psi^{\dag}_{k} = \begin{pmatrix} \overline{c}_{k} & c_{-k} \end{pmatrix}$. It will be useful to work with the fluctuating inverse fermion propagator,
\begin{equation}
    \mathbb{G}^{-1}_{k+q,k}
    =
    \begin{pmatrix}
        (i \omega_n - \xi_{\vb{k}}) \delta_{q, 0} 
        & \Delta_{\vb{k},q} \\
        \overline{\Delta}_{\vb{k},q} 
        & (i \omega_n + \xi_{-\vb{k}}) \delta_{q, 0}
    \end{pmatrix},
\end{equation}
where $\Delta_{\vb{k}, q} = \sum_{\ell} \Delta^{(\ell)}_q \chi^{(\ell)}_{\vb{k}}$.

\subsection{4. Mean-field theory}
Integrating the fermion out from Eq.~\eqref{eq:HS_action} is trivial,
\begin{equation}
    \int D[\Psi]
    \exp(\frac{\beta}{2} \Psi^{\dag} \mathbb{G}^{-1} \Psi)
    =
    \Det \left( - \frac{\beta}{2} \mathbb{G}^{-1} \right)
    =
    \exp[\log \Det \left( - \frac{\beta}{2} \mathbb{G}^{-1} \right)],
\end{equation}
so the exact action for the Hubbard-Stratonovich bosons is
\begin{equation}    \label{eq:HS_action_integrated_out}
    S[\Delta]
    =
    L^2 \beta \sum_{q \ell}
    \frac{|\Delta^{(\ell)}_q|^2}{g_{\ell}}
    - \log \Det \left( -\frac{\beta}{2} \mathbb{G}^{-1} \right)
    =
    L^2 \beta \sum_{q \ell}
    \frac{|\Delta^{(\ell)}_q|^2}{g_{\ell}}
    - \Tr \log(- \frac{\beta}{2} \mathbb{G}^{-1})
\end{equation}
where the capital-D $\Det$ and capital-T $\Tr$ run over spacetime-momentum indices as well as Nambu space. The mean-field ground state corresponds to the saddle point of the action in Eq.~\eqref{eq:HS_action_integrated_out}. Assuming a spatially-and-temporally-uniform configuration $\Delta^{(\ell)}_q = \Deltamf^{(\ell)} \delta_{q,0}$,
\begin{equation}
    S [\Delta]
    \rightarrow
    L^2 \beta \sum_{\ell}
    \frac{|\Deltamf^{(\ell)}|^2}{g_{\ell}}
    -
    \sumprime_k \tr \log(
        -\frac{\beta}{2} \sum_{\ell}
        \begin{pmatrix}
            i \omega_n - \xi_{\vb{k}}
            & \Delta_{\text{mf}}^{(\ell)} \chi^{(\ell)}_{\vb{k}} \\
            \overline{\Delta}_{\text{mf}}^{(\ell)} \overline{\chi}^{(\ell)}_{\vb{k}}
            & i \omega_n + \xi_{-\vb{k}}
        \end{pmatrix}
    ).
\end{equation}
The saddle-point condition then becomes
\begin{equation}
    \frac{\delta S [\Delta]}{\delta \overline{\Delta}_{\text{mf}}^{(\ell)}}
    \rightarrow
    L^2 \beta \frac{\Delta^{(\ell)}_{\text{mf}}}{g_{\ell}}
    -
    \frac{\partial}{\partial \overline{\Delta}^{(\ell)}_{\text{mf}}}
    \sumprime_k \tr \log \left[
        - \frac{\beta}{2}
        \sum_{\ell'}
        \begin{pmatrix}
            i \omega_n - \xi_{\vb{k}}
            & \Delta_{\text{mf}}^{(\ell')} \chi^{(\ell')}_{\vb{k}} \\
            \overline{\Delta}_{\text{mf}}^{(\ell')} \overline{\chi}^{(\ell')}_{\vb{k}}
            & i \omega_n + \xi_{-\vb{k}}
        \end{pmatrix}
    \right]
    =
    0,
\end{equation}
so
\begin{equation}
    \Deltamf^{(\ell)}
    =
    \frac{g_{\ell}}{L^2 \beta} \sumprime_k \tr[
        \begin{pmatrix}
            i \omega_n - \xi_{\vb{k}}
            & \sum_{\ell'} \Delta_{\text{mf}}^{(\ell')} \chi^{(\ell')}_{\vb{k}} \\
            \sum_{\ell'} \overline{\Delta}_{\text{mf}}^{(\ell')} \overline{\chi}^{(\ell')}_{\vb{k}}
            & i \omega_n + \xi_{-\vb{k}}
        \end{pmatrix}^{-1}
        \begin{pmatrix}
            0 & 0 \\
            \overline{\chi}^{(\ell)}_{\vb{k}} & 0
        \end{pmatrix}
    ].
\end{equation}
In general the mean-field gap function has contributions from each Hubbard-Stratonovich (HS) boson: $\Delta_{\vb{k}} = \sum_{\ell} \Deltamf^{(\ell)} \chi^{(\ell)}_{\vb{k}}$. Working at $T \rightarrow 0$ and $L \rightarrow \infty$, and defining the effective inversion-even quasiparticle energy $E_{0,\vb{k}} = \sqrt{\xi_{0,\vb{k}}^2 + |\Delta_{\vb{k}}|^2}$, we have
\begin{equation}    \label{eq:HS_saddle_eqs}
    \Deltamf^{(\ell)}
    =
    g_{\ell}
    \int' \frac{\dd^2\vb{k'}}{(2\pi)^2}
    \int \frac{\dd{\omega}}{2 \pi}
    \frac{\overline{\chi}^{(\ell)}_{\vb{k}'} \Delta_{\vb{k'}}}{
        (i \omega - \tilde{\xi}_{\vb{k'}})^2 - E_{0,\vb{k}}^2
    }
    =
    g_{\ell}
    \int' \frac{\dd^2\vb{k'}}{(2\pi)^2}
    \frac{\overline{\chi}^{(\ell)}_{\vb{k}'} \Delta_{\vb{k'}}}{2 E_{0,\vb{k'}}}
    \Theta(E_{0,\vb{k'}} - |\tilde{\xi}_{\vb{k'}}|),
\end{equation}
where $\Theta$ is the Heaviside step function.
The gap function itself satisfies
\begin{equation}    \label{eq:warped_gap_equation}
    \Delta_{\vb{k}}
    =
    -\int' \frac{\dd^2\vb{k'}}{(2\pi)^2} V_{\vb{k k'}} 
    \frac{\Delta_{\vb{k'}}}{E_{0,\vb{k'}}}
    \Theta(E_{0,\vb{k'}} - |\tilde{\xi}_{\vb{k'}}|).
\end{equation}
The second equality in Eq.~\eqref{eq:HS_saddle_eqs} may be obtained by evaluating the $\omega$ integral as a contour integral in the complex plane. The integrand has poles at $\omega = -i \tilde{\xi}_{\vb{k'}} \pm i E_{0,\vb{k'}}$, whose residues differ only by a minus sign. When $|\tilde{\xi}_{\vb{k'}}| < E_{0,\vb{k'}}$ one pole lives in the upper half-plane, and the other in the lower. We close the integration contour at $\pm \infty$ by a contour enclosing either half-plane, picking up one pole. The reversed sign of the residue between the two poles is compensated by the reversed direction of integration. On the other hand, when $|\tilde{\xi}_{\vb{k'}}| > E_{0,\vb{k'}}$, the poles both lie in the same half-plane. Again, we can close the contour in either direction: either the half-plane devoid of poles delivers zero, or the two residues cancel.

The vanishing of the momentum integrand when $|\tilde{\xi}_{\vb{k'}}| > E_{0,\vb{k'}}$ reflects the fact that, by breaking the degeneracy $\xi_{\vb{k}} = \xi_{-\vb{k}}$ in the free-particle theory, sufficiently strong inversion breaking $\tilde{\xi}$ will completely suppress pairing between those $\vb{k}$ and $-\vb{k}$; without degeneracy between time-reversed states, superconductivity with all Cooper pairs at the same momentum is no longer a weak-coupling instability. We will assume throughout the remainder of this work that the inversion breaking $\tilde{\xi}$ is sufficiently weak compared to pairing interactions and inversion-even higher-order terms in the dispersion (like $\alpha |\vb{k}|^4$) that the condition $|\tilde{\xi}_{\vb{k}}| < E_{0,\vb{k}}$ is ensured to hold for all $\vb{k}$. Where $\xi_{0}$ is small, the condition $|\tilde{\xi}| < E_0$ may be achieved by pairing strong enough to keep the gap $|\Delta| \sim E_0$ self-consistently larger than $|\tilde{\xi}|$. Where $\xi_0$ is large in magnitude (and therefore $E_0 \sim |\xi_0|$), it can do the job itself. With the above assumptions in mind, we arrive at a more familiar gap equation:
\begin{equation}    \label{eq:simple_gap_equation}
    \Delta_{\vb{k}}
    =
    -\int' \frac{\dd^2\vb{k'}}{(2\pi)^2}
    V_{\vb{k k'}} \frac{\Delta_{\vb{k'}}}{E_{0,\vb{k'}}}.
\end{equation}
Note the factor of $2$ on the right-hand-side relative to the usual form, stemming from the $1/2$ in $V_{\vb{kk'}} = \frac{1}{2} \sum_{\ell} g_{\ell} \chi^{(\ell)}_{\vb{k}} \overline{\chi}^{(\ell)}_{\vb{k'}}$.

For our simple model, we can readily verify the chiral solutions $\Delta_{\vb{k}} = \Deltamf e^{\pm i \varphi_{\vb{k}}}$:
\begin{equation}
    \Deltamf e^{\pm i \varphi}
    = \int' \frac{\dd{k'} k'}{2 \pi}
    \int_0^{2\pi} \frac{\dd{\varphi'}}{2\pi}
    \frac{
        g_1 e^{i(\varphi - \varphi')} 
        + g_2 e^{-i(\varphi - \varphi')}
    }{2}
    \frac{
        \Deltamf e^{\pm i \varphi'}
    }{\sqrt{\xi_{0,\vb{k'}}^2 + \Deltamf^2}}
    \\ = 
    g_{1/2}
    \int' \frac{\dd{k'} k'}{2 \pi}
    \frac{\Deltamf e^{\pm i \varphi}}{
        2 \sqrt{\xi_{0,\vb{k'}}^2 + \Deltamf^2}
    },
\end{equation}
i.e.,
\begin{equation}
    \frac{1}{g_{1/2}}
    =
    \int' \frac{\dd{k} k}{2\pi}
    \frac{1}{2 \sqrt{\xi_{0, k}^2 + \Deltamf^2}},
\end{equation}
where the choice of $1/2$ goes with the choice of $+/-$. 
We can approximate the solution as usual, by going over to an energy integral via the density of states $\nu$:
\begin{equation}
    \int' \frac{\dd{k} k}{2 \pi}
    \rightarrow \nu \int_{-\Lambda}^{\Lambda} \dd{\xi},
\end{equation}
yielding the familiar result
\begin{equation}    \label{eq:gap_result}
    \frac{1}{g_{1/2}}
    \approx
    \nu \int_{-\Lambda}^{\Lambda} \dd{\xi}
    \frac{1}{2 \sqrt{\xi^2 + |\Deltamf|^2}}
    =
    \nu \arctanh \frac{\Lambda}{\sqrt{\Lambda^2 + |\Deltamf|^2}}
    \approx \nu \log \frac{2 \Lambda}{|\Deltamf|}
\end{equation}
where the final approximation holds for $\Lambda \gg \Deltamf$. Clearly, then, $\Delta_{\vb{k}} = \Deltamf e^{\pm i \varphi}$ are both self-consistent solutions. Choosing $g_1 > g_2$ means that the $p + ip$ solution has a larger gap $\Deltamf$ than does the $p-ip$ solution, hence it saves more condensation energy, and wins the competition for ground state. We then have
\begin{equation}    \label{eq:inverse_gs}
    \frac{1}{g_1} = \nu \log \frac{2 \Lambda}{\Deltamf}
    \qquad \text{and} \qquad
    \frac{1}{g_2}
    =
    \frac{1}{g_1 - \delta g}
    =
    \nu \log(\frac{2 \Lambda}{\Deltamf})
    \left( 1 + \frac{\delta g / g_1}{1 - \delta g / g_1} \right).
\end{equation}

\subsection{5. The electromagnetic vertices}
We now couple to electromagnetism, introducing the gauge field $A = (A_0, \vb{A})$ as
\begin{equation}
\label{app:eq_Peierls}
    \overline{\psi} \left[
        \partial_{\tau} 
        + \xi\left( \frac{1}{i} \grad \right)
    \right] \psi
    \rightarrow
    \overline{\psi} \left[
        \partial_{\tau} + i e A_0
        + \xi\left( \frac{1}{i} \grad - e \vb{A} \right)
    \right] \psi.
\end{equation}
To find the vector-potential vertices, note that the kinetic energy term generally involves spatial derivatives of both the vector potential and the fermion. We can define the operator $\xi(\frac{1}{i} \grad - e \vb{A})$ by its series expansion,
\begin{equation}
    \xi\left( \frac{1}{i} \grad - e \vb{A} \right)
    =
    \xi(0)
    + \frac{1}{i} \xi'_j(0) \left( 
        \partial_j - i e A_j
    \right)
    + \frac{1}{i^2} \frac{1}{2!} \xi''_{j_1 j_2}(0) \left( 
        \partial_{j_1} - i e A_{j_1}
    \right) \left( 
        \partial_{j_2} - i e A_{j_2}
    \right)
    + \dots.
\end{equation}
The $O(\vb{A}^0)$ piece is the free-particle dispersion $\xi(-i\grad)$. Dropping the explicit argument $(0)$ on each derivative of $\xi$ for brevity, we obtain the paramagnetic term as the $O(\vb{A})$ piece,
\begin{multline}
    -i e \left[
        \frac{1}{i} \xi'_j A_j
        +
        \frac{1}{i^2} 
        \frac{1}{2!}
        \xi''_{j_1 j_2} \left(
            A_{j_1} \partial_{j_2}
            + \partial_{j_1} A_{j_2}
        \right)
        +
        \frac{1}{i^3}
        \frac{1}{3!} \xi'''_{j_1 j_2 j_3} \left(
            A_{j_1} \partial_{j_2} \partial_{j_3}
            + \partial_{j_1} A_{j_2} \partial_{j_3}
            + \partial_{j_1} \partial_{j_2} A_{j_3}
        \right)
        + 
        \dots
    \right]
    \\
    =
    -e \sum_{m=1}^{\infty}
    \frac{1}{i^{m-1}}
    \frac{1}{m!} 
    \xi^{(m)}_{j_1 \dots j_m}
    \sum_{n=0}^{m-1}
    \partial^n_{j_1 \dots j_n} A_{j_{n+1}}
    \partial^{m-1-n}_{j_{n+2} \dots j_m},
\end{multline}
and the diamagnetic term as the $O(\vb{A}^2)$ piece,
\begin{multline}
    -e^2 \left[
        \frac{1}{i^2} \frac{1}{2!} \xi''_{j_1 j_2} A_{j_1} A_{j_2}
        + \frac{1}{i^3} \frac{1}{3!} \xi'''_{j_1 j_2 j_3} \left(
            A_{j_1} A_{j_2} \partial_{j_3}
            + A_{j_1} \partial_{j_2} A_{j_3}
            + \partial_{j_1} A_{j_2} A_{j_3}
        \right)
        + \dots
    \right]
    \\
    =
    -e^2 \sum_{m=2}^{\infty}
    \frac{1}{i^m} \frac{1}{m!} \xi^{(m)}_{j_1 \dots j_m}
    \sum_{n=0}^{m-2}
    \sum_{\ell=0}^{m-2-\ell}
    \partial^n_{j_1 \dots j_n} A_{j_{n+1}} \partial^{\ell}_{j_{n+2} \dots j_{n+\ell+1}} A_{j_{n+\ell+2}} \partial^{m-n-\ell-2}_{j_{n+\ell+3} \dots j_m}.
\end{multline}
These expressions are usually rather impractical to work with since $[\partial_i, A_j] = \partial_i(A_j) \ne 0$; they simplify for a parabolic band, but not in the general case. In the spatially-uniform limit $\vb{q}\rightarrow0$ these terms simplify drastically since $[\partial_i, A_j] = 0$, so we can work directly in momentum space and Taylor expand around $\vb{k}$:
\begin{equation}
\label{app:eq_Taylor}
    \xi_{\vb{k} - e\vb{A}}
    =
    \xi_{\vb{k}} + \xi'_{j_1, \vb{k}} (-e A_{j_1})
    + \frac{1}{2!} \xi''_{j_1 j_2, \vb{k}} (-e A_{j_1}) (-e A_{j_2})
    + \dots.
\end{equation}

Going to the Nambu representation $\Psi_k = \begin{pmatrix} c_k & \overline{c}_{-k} \end{pmatrix}^T$, the scalar-potential vertex is immediately obtained from Eq.~\eqref{app:eq_Peierls}:
\begin{equation}
    \mathbb{V}^{(\Ay_0)}_{k+q, k}
    =
    i e \Ay_{0,q} \begin{pmatrix}
        1 & 0 \\
        0 & -1
    \end{pmatrix}.
\end{equation}
We then read off the vector-potential vertices from Eq.~\eqref{app:eq_Taylor} ($q$ is here an alias for $\Omega_m$ since we took $\vb{q}=0$):
\begin{equation}
    \mathbb{V}^{(\bAy)}_{k+q, k}
    =
    - e \bAy_q \vdot \begin{pmatrix}
        \vb{v}_{\vb{k}} & 0 \\
        0 & -\vb{v}_{\vb{-k}}
    \end{pmatrix}
\end{equation}
and
\begin{equation}
    \mathbb{V}^{(\bAy^2)}_{k+q, k}
    =
    \frac{e^2}{2} \sum_p
    \Ay_{i,q-p} \Ay_{j,p}
    \begin{pmatrix}
        (m^{-1}_{\vb{k}})_{ij} & 0 \\
        0 & -(m^{-1}_{-\vb{k}})_{ij}
    \end{pmatrix}.
\end{equation}
$(m^{-1}_{\vb{k}})_{ij} = \partial_{k_i} v^j_{\vb{k}} = \partial_{k_i} \partial_{k_j} \xi_{\vb{k}}$ is the inverse effective mass tensor. We wrote $\Ay$ instead of $\vb{A}$ since gauging out the phase as we do turns all $A$s into $\Ay$s. Alternatively, one can simply invoke gauge invariance to see that before integrating over $\theta$, $A$ and $\theta$ can only appear in the effective action via the combination $\Ay$.

\subsection{6. Fermionic contribution to the response tensor}
\subsubsection{a) Superfluid stiffness}
Before considering the impact of collective modes, the ``bare'' superfluid stiffness of the condensate can be obtained from a textbook calculation~\cite{coleman2015manybody}, which we present here generalized slightly to account for broken inversion. There are two bubble diagrams: one with a single diamagnetic vertex, and the other with two paramagnetic vertices,
\begin{equation}
    \sum_q
    K^{i j}_{\text{f}, q} \Ay_{j,-q} \Ay_{j, q}
    =
    \frac{1}{L^2 \beta} \sum_k \tr{
        \mathcal{G}_k
        \mathbb{V}^{(\bAy^2)}_{k k}
    }
    +
    \frac{1}{2} \frac{1}{L^2 \beta} 
    \sum_{kq} \tr{
        \mathcal{G}_{k+q} 
        \mathbb{V}^{(\bAy)}_{k+q, k}
        \mathcal{G}_{k}
        \mathbb{V}^{(\bAy)}_{k, k+q}
    }.
\end{equation}
We can turn the diamagnetic term into a two-vertex bubble by integrating by parts over the spatial momentum and using $\partial_{k_i} \mathcal{G}_k = - \mathcal{G}_k (\partial_{k_i} \mathcal{G}^{-1}_k) \mathcal{G}_k$:
\begin{multline}
    \frac{1}{\beta} \sum_{i \omega_n} 
    \int \frac{\dd^2{\vb{k}}}{(2 \pi)^2}
    \tr{
        \mathcal{G}_k 
        \mathbb{V}^{(\bAy^2)}_{k k}
    }
    =
    \sum_q
    \frac{e^2}{2} \frac{1}{\beta} \sum_{i \omega_n} 
    \Ay_{i,-q} \Ay_{j,q}
    \int \frac{\dd^2{\vb{k}}}{(2 \pi)^2}
    \tr{
        \mathcal{G}_k
        \begin{pmatrix}
            \partial_{k_i} \partial_{k_j}
            \xi_{\vb{k}} & 0 \\
            0 & -\partial_{k_i} \partial_{k_j}
            \xi_{-\vb{k}}
        \end{pmatrix}
    }
    \\ =
    -\sum_q
    \frac{e^2}{2} \frac{1}{\beta} \sum_{i \omega_n} 
    \Ay_{i,-q} \Ay_{j,q}
    \int \frac{\dd^2{\vb{k}}}{(2 \pi)^2}
    \tr{
        \partial_{k_i} \mathcal{G}_k
        \begin{pmatrix}
            v^j_{\vb{k}} & 0 \\
            0 & v^j_{-\vb{k}}
        \end{pmatrix}
    }
    \\ =
    \sum_q
    \frac{e^2}{2} \frac{1}{\beta} \sum_{i \omega_n} 
    \Ay_{i,-q} \Ay_{j,q}
    \int \frac{\dd^2{\vb{k}}}{(2 \pi)^2}
    \tr{
        \mathcal{G}_k (\partial_{k_i} \mathcal{G}^{-1}_k) \mathcal{G}_k
        \begin{pmatrix}
            v^j_{\vb{k}} & 0 \\
            0 & v^j_{-\vb{k}}
        \end{pmatrix}
    }
    \\=
    -\sum_q
    \frac{e^2}{2} \frac{1}{\beta} \sum_{i \omega_n} 
    \Ay_{i,-q} \Ay_{j,q}
    \int \frac{\dd^2{\vb{k}}}{(2 \pi)^2}
    \tr{
        \mathcal{G}_k
        \begin{pmatrix}
            v^i_{\vb{k}} & -\partial_{k_i} \Delta_{\vb{k}} \\
            -\partial_{k_i} \overline{\Delta}_{\vb{k}}
            & v^i_{-\vb{k}}
        \end{pmatrix}
        \mathcal{G}_k
        \begin{pmatrix}
            v^j_{\vb{k}} & 0 \\
            0 & v^j_{-\vb{k}}
        \end{pmatrix}
    }
\end{multline}
where summation on $i$ and $j$ is again implicit. Combining the paramagnetic and diamagnetic contributions,
\begin{multline}    \label{eq:fermionic_stiffness}
    K^{i j}_{\text{f}, \vb{q}=0}(i\Omega_m)
    \\
    =
    \frac{e^2}{2} \frac{1}{\beta} \sum_{i \omega_n}
    \int \frac{\dd^2{\vb{k}}}{(2 \pi)^2}
    \tr[
        \mathcal{G}_{k_-}
        \begin{pmatrix}
            v^i_{\vb{k}} & 0 \\
            0 & -v^i_{-\vb{k}}
        \end{pmatrix}
        \mathcal{G}_{k_+}
        \begin{pmatrix}
            v^j_{\vb{k}} & 0 \\
            0 & -v^j_{-\vb{k}}
        \end{pmatrix}
        -
        \mathcal{G}_k
        \begin{pmatrix}
            v^i_{\vb{k}} & -\partial_{k_i} \Delta_{\vb{k}} \\
            -\partial_{k_i} \overline{\Delta}_{\vb{k}}
            & v^i_{-\vb{k}}
        \end{pmatrix}
        \mathcal{G}_k
        \begin{pmatrix}
            v^j_{\vb{k}} & 0 \\
            0 & v^j_{-\vb{k}}
        \end{pmatrix}
    ]
    \\ \xrightarrow{T \rightarrow 0}
    \frac{e^2}{2}
    \int \frac{\dd^2{\vb{k}}}{(2 \pi)^2}
    \left[
            |\Delta_{\vb{k}}|^2 \left(
                \frac{v^i_{0,\vb{k}} v^j_{0,\vb{k}}}{E_{0, \vb{k}}^3}
                +
                \frac{
                    4 \tilde{v}^i_{\vb{k}} \tilde{v}^j_{\vb{k}}
                }{
                    E_{0,\vb{k}} 
                    \left[ (i \Omega_m)^2 - 4 E_{0,\vb{k}}^2 \right]
                }
            \right)
        -
        \frac{\xi_{0,\vb{k}} v^j_{0, \vb{k}}}{E_{0,\vb{k}}^3} 
        \Re \left(
            \overline{\Delta}_{\vb{k}}
            \partial_{k_i} \Delta_{\vb{k}}
        \right)
    \right]
    \\
    =
    \frac{e^2}{2}
    \int \frac{\dd^2{\vb{k}}}{(2 \pi)^2}
    \frac{1}{E_{0, \vb{k}}^3}
    \left[
        |\Delta_{\vb{k}}|^2 \left(
            v^i_{0,\vb{k}} v^j_{0,\vb{k}}
            +
            \tilde{v}^i_{\vb{k}} \tilde{v}^j_{\vb{k}}
            \frac{4 E_{0, \vb{k}}^2}{(i \Omega_m)^2 - 4 E_{0, \vb{k}}^2}
        \right)
        -
        \frac{1}{2} \xi_{0,\vb{k}}
        (\partial_{k_i} |\Delta_{\vb{k}}|^2) v^j_{0, \vb{k}}
    \right].
\end{multline}
Here we introduced $k_{\pm} = k \pm \frac{1}{2} q$ (which is only a valid $(d+1)$-momentum in the $L \rightarrow \infty$, $T \rightarrow 0$ limit). Writing an arbitrary gap function as $\Delta_{\vb{k}} = |\Delta_{\vb{k}}| e^{i \theta_{\vb{k}}}$, the last line used
\begin{equation}
    \Re (\overline{\Delta} \partial \Delta)
    =
    \Re \left[ |\Delta| e^{-i \theta} \partial (e^{i \theta} |\Delta|) \right] 
    =
    |\Delta| \partial |\Delta|
    =
    \frac{1}{2} \partial |\Delta|^2
\end{equation}
where obvious subscripts are suppressed for brevity. In our simple model where $\Delta_{\vb{k}} = |\Delta| e^{i \varphi_{\vb{k}}}$, the last term $\propto \partial_{k_i} |\Delta_{\vb{k}}|^2$ vanishes. Note that despite its simple appearance, Eq.~\eqref{eq:fermionic_stiffness} makes no approximation with respect to warping (after the assumption that $|\tilde{\xi}| < E_0$ everywhere). Surprisingly, the second-order correction $\propto \tilde{v}^2$ is exact. 

In our simple model, the result simplifies further. First,
\begin{multline}
    K^{ij}_{\text{f}, \vb{q}=0}(i\Omega_m)
    =
    \frac{e^2}{2}
    \int \frac{\dd^2{\vb{k}}}{(2 \pi)^2}
    \frac{\Deltamf^2}{E_{0, \vb{k}}^3}
    \left[
        v^i_{0,\vb{k}} v^j_{0,\vb{k}}
        +
        \tilde{v}^i_{\vb{k}} \tilde{v}^j_{\vb{k}}
        \frac{4 E_{0, \vb{k}}^2}{(i \Omega_m)^2 - 4 E_{0, \vb{k}}^2}
    \right]
    \\
    =
    \delta_{ij} \frac{e^2}{4}
    \int \frac{\dd{k} k}{2 \pi}
    \frac{\Deltamf^2}{E_{0, \vb{k}}^3}
    \left[
        \frac{k^2}{m^2}
        +
        (\eta k^2)^2
        \frac{4 E_{0, \vb{k}}^2}{(i \Omega_m)^2 - 4 E_{0, \vb{k}}^2}
    \right].
\end{multline}
Going over to an energy integral via the density of states, and separating convergent and divergent parts of the integral,
\begin{multline}
    K^{ij}_{\text{f}, \vb{q}=0} (i \Omega_m)
    =
    \delta_{ij} e^2 \nu
    \int \dd{\xi} \Bigg[
        \frac{\Deltamf^2}{(\xi^2 + \Deltamf^2)^{3/2}} \frac{\xi + \mu}{2 m}
        +
        \frac{4 (m \eta \Deltamf)^2 (\xi + \mu)^2}{\sqrt{\xi^2 + \Deltamf^2}}
        \frac{1}{(i \Omega_m)^2 - 4 \xi^2 - 4 \Deltamf^2}
    \Bigg]
    \\ \\ =
    \delta_{ij} e^2 \nu \int_{-\infty}^{\infty} \dd{\xi} \left[
        \frac{\Deltamf^2}{(\xi^2 + \Deltamf^2)^{3/2}} \frac{\mu}{2 m}
        + \frac{4 (m \eta \mu)^2}{\sqrt{\xi^2 + \Deltamf^2}} \frac{\Deltamf^2}{(i \Omega_m)^2 - 4 \xi^2 - 4 \Deltamf^2}
    \right]
    \\ + \delta_{ij} e^2 \nu \int_{-\Lambda}^{\Lambda} \dd{\xi}
    \frac{4 (m \eta \Deltamf)^2}{\sqrt{\xi^2 + \Deltamf^2}} \frac{\xi^2}{(i \Omega_m)^2 - 4 \xi^2 - 4 \Deltamf^2}
    \\ \\ =
    \delta_{ij} e^2 \nu \Bigg[
        \frac{\mu}{m} - 2 (\eta m \mu)^2 \frac{4 \Deltamf^2}{i |\Omega_m| \sqrt{4 \Deltamf^2 - (i \Omega_m)^2}}
        \arcsec \frac{2 \Deltamf}{\sqrt{4 \Deltamf^2 - (i \Omega_m)^2}}
        \\
        + 2 (\eta m \Deltamf)^2
        \frac{\sqrt{4 \Deltamf^2 - (i \Omega_m)^2}}{\Omega_m}
        \arctanh \frac{\Lambda \Omega_m}{\sqrt{(\Deltamf^2 + \Lambda^2)[4 \Deltamf^2 - (i \Omega_m)^2]}}
        \\
        - 2 (\eta m \Deltamf)^2 \arctanh \frac{\Lambda}{\sqrt{\Deltamf^2 + \Lambda^2}}
    \Bigg].
\end{multline}
Taking the large-cutoff limit $\Lambda \gg \Deltamf$ gives
\begin{multline}
    K^{ij}_{\text{f}, \vb{q}=0} \approx
    \delta_{ij} e^2 \nu \Bigg[
        \frac{\mu}{m} - 2 (\eta m \mu)^2 \frac{(2 \Deltamf)^2}{i |\Omega_m| \sqrt{4 \Deltamf^2 - (i \Omega_m)^2}}
        \arcsin \frac{i \Omega_m}{2 \Deltamf} 
        \\ + 2 (\eta m \Deltamf)^2
        \frac{\sqrt{4 \Deltamf^2 - (i \Omega_m)^2}}{i \Omega_m}
        \arcsin \frac{i \Omega_m}{2 \Deltamf}
        -2 (\eta m \Deltamf)^2 \log \frac{2 \Lambda}{\Deltamf}
    \Bigg].
\end{multline}
Analytically continuing to real time, $i \Omega_m \rightarrow \Omega + i 0^+$, we have
\begin{equation}    \label{eq:realtime_fermion_stiffness}
    K^{ij}_{\text{f},\vb{q}=0} (\Omega)
    =
    \delta_{ij} e^2 \nu \left\lbrace
        \frac{\mu}{m}
        - 2 (\eta m \mu)^2 \left[
            z \sec z \csc z
            - \left(\frac{\Deltamf}{\mu}\right)^2
            \left( z  \cot z - \log \frac{2 \Lambda}{\Deltamf} \right)
        \right]
    \right\rbrace
\end{equation}
where the requirement of analyticity on the upper half-plane lets us replace $|\Omega_m| \rightarrow \Omega_m \rightarrow -i \Omega + 0^+$ in the denominator, and where
\begin{equation}
    \sin z = \frac{\Omega + i 0^+}{2 \Deltamf}.
\end{equation}

\subsubsection{b) Compressibility}
The bare-fermionic contribution to the compressibility corresponds to
\begin{equation}
    K^{00}_{\text{f}, q} \Ay_{0,-q} \Ay_{0,q}
    =
    \frac{1}{2} \frac{1}{L^2\beta} \sum_k \tr{
        \mathcal{G}_{k} \mathbb{V}^{(\Ay_0)}_{k,k+q}
        \mathcal{G}_{k+q} \mathbb{V}^{(\Ay_0)}_{k+q,k}
    },
    \quad \text{i.e.,} \quad
    K^{00}_{\text{f}, q}
    =
    -\frac{e^2}{2 L^2 \beta} \sum_k \tr{
        \mathcal{G}_k \tau^z \mathcal{G}_{k+q} \tau^z
    }.
\end{equation}
One finds (again assuming $|\tilde{\xi}| < E_0$)
\begin{multline}
    K^{00}_{\text{f}, \vb{q}=0}(i\Omega_m)
    \xrightarrow[L\rightarrow\infty]{T\rightarrow0}
    -\frac{e^2}{2} \int \frac{\dd^2{\vb{k}}}{(2\pi)^2}
    \frac{1}{E_{0,\vb{k}}}
    \frac{|2 \Delta_{\vb{k}}|^2}{(i \Omega_m)^2 - 4 E_{0,\vb{k}}^2}
    =
    -e^2 \nu \int_{-\infty}^{\infty} \dd{\xi}
    \frac{1}{E}
    \frac{2 \Deltamf^2}{(i \Omega_m)^2 - 4 E^2}
    \\ =
    \sgn(\Omega_m) \frac{
        4 \Deltamf^2 \arcsec\left( 
            \frac{2 \Deltamf}{\sqrt{4\Deltamf^2 - (i\Omega_m)^2}}
        \right)
    }{
        i \Omega_m \sqrt{4\Deltamf^2 - (i\Omega_m)^2}
    }
\end{multline}
where we used $E = \sqrt{\xi^2 + \Deltamf^2}$.

\subsection{7. Collective-mode contributions to the superfluid stiffness}
\subsubsection{a) The collective-mode propagator}
The collective-mode propagator inverts the combined Hubbard-Stratonovich $1/g$ term and generalized polarization bubbles $M^{\phi \phi'}$ (with $\phi, \phi' \in \lbrace h, a, b \rbrace$):
\begin{equation}
    -\Phi^T_{-q} \hat{\mathcal{D}}^{-1}_{q} \Phi_q
    =
    \Phi^T_{-q}
    \begin{pmatrix}
        \frac{1}{g_1} & 0 & 0 \\
        0 & \frac{1}{g_2} & 0 \\
        0 & 0 & \frac{1}{g_2}
    \end{pmatrix}
    \Phi_q
    + \underbrace{\frac{1}{2} \frac{1}{L^2 \beta} \sumprime_{k} \tr{
        \mathcal{G}_{k+q} \mathbb{V}^{(\Phi)}_{k+q,k}
        \mathcal{G}_{k} \mathbb{V}^{(\Phi)}_{k,k+q}
    }}_{\Phi^T_{-q} \hat{M}_q \Phi_q}
\end{equation}
where we lumped together the pair-fluctuation vertices into $\mathbb{V}^{(\Phi)} = \mathbb{V}^{(h)} + \mathbb{V}^{(a)} + \mathbb{V}^{(b)}$. One finds readily (recall $\Delta_{\vb{k}} = \Deltamf \chi^{(1)}_{\vb{k}}$)
\begin{multline}
    \hat{M}_{\vb{q}=0}(i\Omega_m)
    \xrightarrow{T\rightarrow0}
    2 \int' \frac{\dd^2\vb{k}}{(2 \pi)^2}
    \frac{1}{E_{0,\vb{k}}}
    \frac{1}{(i \Omega_m)^2 - 4 E_{0, \vb{k}}^2}
    \\
    \begin{pmatrix}
        \xi_{0, \vb{k}}^2 \left|\chi^{(1)}_{\vb{k}}\right|^2
        & \xi_{0, \vb{k}} \Re \left[
            \overline{\chi}^{(1)}_{\vb{k}} \chi^{(2)}_{\vb{k}} (\xi_{0,\vb{k}} - \frac{1}{2} i \Omega_m)
        \right]
        & -\xi_{0,\vb{k}} \Im \left[
            \overline{\chi}^{(1)}_{\vb{k}} \chi^{(2)}_{\vb{k}} (\xi_{0,\vb{k}} - \frac{1}{2} i \Omega_m)
        \right]
        \\ \cdot
        & \xi_{0,\vb{k}}^2 \left| \chi^{(2)}_{\vb{k}} \right|^2
        + \Deltamf^2 \Im \left[ \chi^{(1)}_{\vb{k}} \overline{\chi}^{(2)}_{\vb{k}} \right]^2
        & \frac{1}{2} \xi_{0,\vb{k}} \Omega_m \left| \chi^{(2)}_{\vb{k}} \right|^2
        - \frac{1}{2}  \Deltamf^2 \Im \left[
            \left( \chi^{(1)}_{\vb{k}} \right)^2 \left( \overline{\chi}^{(2)}_{\vb{k}} \right)^2
        \right]
        \\ \cdot
        & \cdot
        & \xi_{0,\vb{k}}^2 \left| \chi^{(2)}_{\vb{k}} \right|^2
        + \Deltamf^2 \Re \left[ \chi^{(1)}_{\vb{k}} \overline{\chi}^{(2)}_{\vb{k}} \right]^2
    \end{pmatrix}
\end{multline}
where the entries marked by dots ($\cdot$) are obtained from $M^{\phi \phi'}_{\vb{q}=0}(i\Omega_m) = M^{\phi' \phi}_{\vb{q}=0}(-i\Omega_m)$. Note that at $T \rightarrow 0$, all of these $\omega$ integrals instead vanish when $|\tilde{\xi}| > \sqrt{\xi_0^2 + |\Delta|^2}$; the situation is the same as for the mean-field equation discussed in the previous section, and can be understood by similar reasoning about contour integrals. In that section we discussed and assumed sufficient conditions to ensure $|\tilde{\xi}_{\vb{k}}| < E_{0,\vb{k}}$ everywhere avoiding complication.

Rotational symmetries will frequently enforce vanishing of the couplings between the Higgs ($h$) and BaSch/clapping modes ($a$ and $b$). For example, consider the $h$-$a$ coupling $\Pi^{ha}$. In the chiral $p$-wave clapping mode case ($\chi^{(1)} = \overline{\chi}^{(2)} = e^{i \varphi}$), the angular integral vanishes assuming $C_{n \ge 4}$ symmetry of $\xi_0$ ($\xi_0$ is even by construction, so has no nontrivial $C_3$ symmetric part). If we consider an $f+if$ BaSch mode over a $p+ip$ ground state, e.g., $\chi^{(1)} = e^{i \varphi}$ and $\chi^{(2)} = e^{3 i \varphi}$, then the $h$-$a$ coupling is permitted if the $\xi_{0,\vb{k}}$ has a nontrivial $C_4$-symmetric component, i.e., a piece $\propto \cos(4 \varphi)$. However, since our main result is that optical couplings arise in trigonally warped systems (incompatible with $C_4$ symmetry), the first interesting case is the $h + i h$ BaSch mode over a $p + ip$ ground state, $\chi^{(2)} = e ^{i 5 \varphi}$. If $\xi_0$ shows hexagonal warping $\propto \cos(6 \varphi)$ then the Higgs and BaSch modes can couple. The same reasoning holds for the clapping mode of an $f + i f$ state with $\chi^{(1)} = \overline{\chi}^{(2)} = e^{3 i \varphi}$.
    
In our simple model of $p+ip$ clapping modes, $\chi^{(1)}_{\vb{k}} = \overline{\chi}^{(2)}_{\vb{k}} = e^{i \varphi_{\vb{k}}}$, and $\xi_{0,\vb{k}}$ has circular symmetry, making the angular integration trivial:
\begin{multline}
    \hat{M}_{\vb{q}=0}(i\Omega_m)
    = \\
    2 \int' \frac{\dd^2\vb{k}}{(2 \pi)^2}
    \frac{1}{E_{0,\vb{k}}}
    \frac{1}{(i \Omega_m)^2 - 4 E_{0, \vb{k}}^2}
    \begin{pmatrix}
        \xi_{0, \vb{k}}^2
        & \xi_{0,\vb{k}}^2 \cos(2\varphi_{\vb{k}})
        + \frac{1}{2} \xi_{0,\vb{k}} \Omega_m  
        \sin(2\varphi_{\vb{k}})
        & \xi_{0,\vb{k}} \left[
            \xi_{0,\vb{k}} \sin(2\varphi_{\vb{k}})
            - \frac{1}{2} \Omega_m \cos(2\varphi_{\vb{k}})
        \right]
        \\ \cdot
        & \xi_{0,\vb{k}}^2 
        + \Deltamf^2 \sin^2(2\varphi_{\vb{k}})
        & -\frac{1}{2} \xi_{0,\vb{k}} \Omega_m 
        - \frac{1}{2}  \Deltamf^2 \sin(4\varphi_{\vb{k}})
        \\ \cdot
        & \cdot
        & \xi_{0,\vb{k}}^2
        + \Deltamf^2 \cos^2(2\varphi_{\vb{k}})
    \end{pmatrix}
    \\
    =
    2 \int' \frac{\dd{k} k}{2 \pi}
    \frac{1}{E_{0,\vb{k}}}
    \frac{1}{(i \Omega_m)^2 - 4 E_{0, \vb{k}}^2}
    \begin{pmatrix}
        \xi_{0, \vb{k}}^2 & 0 & 0 \\
        0 & \xi_{0,\vb{k}}^2  
        + \frac{1}{2} \Deltamf^2
        & -\frac{1}{2} \xi_{0,\vb{k}} \Omega_m 
        \\ 0 & \frac{1}{2} \xi_{0,\vb{k}} \Omega_m 
        & \xi_{0,\vb{k}}^2 + \frac{1}{2} \Deltamf^2
    \end{pmatrix}.
\end{multline}
We then approximate
\begin{multline}
    \hat{M}_{\vb{q}=0}(i \Omega_m)
    \approx
    2 \nu \int_{-\Lambda}^{\Lambda} \dd{\xi}
    \frac{1}{\sqrt{\xi + \Deltamf^2}}
    \frac{1}{(i \Omega_m)^2 - 4 \xi^2 - 4 \Deltamf^2}
    \begin{pmatrix}
        \xi^2 & 0 & 0 \\
        0 & \xi^2 + \frac{1}{2} \Deltamf^2 & 0
        \\ 0 & 0 & \xi^2 + \frac{1}{2} \Deltamf^2
    \end{pmatrix}
    \\ \\ =
    - \nu \arctanh \left( \frac{\Lambda}{\sqrt{\Lambda^2 + \Deltamf^2}} \right) \begin{pmatrix}
        1 & 0 & 0 \\
        0 & 1 & 0 \\
        0 & 0 & 1
    \end{pmatrix}
    \\ +
    \frac{\nu}{\Omega_m} \arctanh \left(
        \frac{\Lambda \Omega_m}{\sqrt{
            (\Lambda^2 + \Deltamf^2)
            (\Omega_m^2 + 4 \Deltamf^2)
        }}
    \right) \begin{pmatrix}
        \sqrt{4 \Deltamf^2 + \Omega_m^2} & 0 & 0 \\
        0 & \frac{2 \Deltamf^2 + \Omega_m^2}{\sqrt{4 \Deltamf^2 + \Omega_m^2}} & 0 \\
        0 & 0 & \frac{2 \Deltamf^2 + \Omega_m^2}{\sqrt{4 \Deltamf^2 + \Omega_m^2}}
    \end{pmatrix}
    \\ \\ \approx
    - \nu \log \frac{2 \Lambda}{\Deltamf}
    \begin{pmatrix}
        1 & 0 & 0 \\
        0 & 1 & 0 \\
        0 & 0 & 1
    \end{pmatrix}
    + \frac{\nu}{i \Omega_m} \arctan \left(
        \frac{i \Omega_m}{
            \sqrt{4 \Deltamf^2 - (i \Omega_m)^2}
        }
    \right) \begin{pmatrix}
        \sqrt{4 \Deltamf^2 - (i \Omega_m)^2} & 0 & 0 \\
        0 & \frac{2 \Deltamf^2 - (i \Omega_m)^2}{\sqrt{4 \Deltamf^2 - (i \Omega_m)^2}} & 0 \\
        0 & 0 & \frac{2 \Deltamf^2 - (i \Omega_m)^2}{\sqrt{4 \Deltamf^2 - (i \Omega_m)^2}}
    \end{pmatrix}.
\end{multline}
For the Higgs mode, the $\log$ divergence cancels exactly against $1/g_1$ via the gap equation \eqref{eq:gap_result}, $1/(g_1 \nu) = \log(2 \Lambda / \Deltamf)$. $\mathcal{T}$ breaking means the cancellation is incomplete for the $a$ and $b$ modes; recall Eq.~\eqref{eq:inverse_gs}
\begin{equation}
    \frac{1}{g_2}
    =
    \nu \log(\frac{2 \Lambda}{\Deltamf})
    \bigg( 1 + \underbrace{\frac{\delta g / g_1}{1 - \delta g / g_1}}_{f} \bigg).
\end{equation}
We are then left with
\begin{multline}    \label{eq:mode_propagator_matrix}
    \hat{\mathcal{D}}^{-1}_{\vb{q}=0}(i\Omega_m)
    =
    - \nu f \log(\frac{2 \Lambda}{\Deltamf})
    \begin{pmatrix}
        0 & 0 & 0 \\
        0 & 1 & 0 \\
        0 & 0 & 1
    \end{pmatrix}
    \\
    -\nu \arctan \left(
        \frac{i \Omega_m}{
            \sqrt{4 \Deltamf^2 - (i \Omega_m)^2}
        }
    \right) \begin{pmatrix}
        \frac{\sqrt{4 \Deltamf^2 - (i \Omega_m)^2}}{i \Omega_m} & 0 & 0 \\
        0 & \frac{2 \Deltamf^2 - (i \Omega_m)^2}{i \Omega_m \sqrt{4 \Deltamf^2 - (i \Omega_m)^2}} & 0 \\
        0 & 0 & \frac{2 \Deltamf^2 - (i \Omega_m)^2}{i \Omega_m \sqrt{4 \Deltamf^2 - (i \Omega_m)^2}}
    \end{pmatrix}.
\end{multline}
After analytic continuation ($i \Omega_m \rightarrow \Omega + i 0^+$), the clapping modes correspond to the zeros of the lower two diagonal components, i.e., the non-Higgs poles of $\hat{\mathcal{D}}$. We read off immediately that those modes remain degenerate after $\mathcal{T}$ breaking, but they are shifted in energy away from $\sqrt{2} \Deltamf$ as one would expect:
\begin{equation}
    \frac{1}{\Omega_c} \arcsin \left(
        \frac{\Omega_c}{ 2 \Deltamf } \right)
    \frac{2 \Deltamf^2 - \Omega_c^2}{\sqrt{4 \Deltamf^2 - \Omega_c^2}}
    +
    \frac{1}{\nu g_1} \frac{\delta g / g_1}{1 - \delta g / g_1}
    =
    0.
\end{equation}
The transcendental function $\arcsin$ obstructs an exact analytical solution away from $\delta g = 0$. We can get some insight into the impact of $\delta g$ by assuming it is small, and expanding $\Omega_c = \sqrt{2}\Deltamf + \delta\Omega_c$:
\begin{equation}
    \frac{\pi}{2} \frac{\delta \Omega_c}{\sqrt{2} \Deltamf} + O(\delta \Omega_c^2)
    =
    \frac{1}{\nu g_1} \frac{\delta g / g_1}{1 - \delta g / g_1},
\end{equation}
i.e.,
\begin{equation}
    \Omega_c
    \approx
    \sqrt{2} \Deltamf \left(
        1 + \frac{2}{\pi}
        \frac{1}{\nu g_1} \frac{\delta g / g_1}{1 - \delta g / g_1}
    \right).
\end{equation}
As expected, lifting the degeneracy between pairing channels raises the clapping mode energy.

\subsubsection{b) Optical couplings}
The couplings to the vector gauge variables are given by Eq.~\eqref{eq:mode_couplings_diagram} of the main text,
\begin{equation}    \label{eq:mode_couplings_supp}
    \Ay_{j,-q} (\Pi^{j}_q)^T \Phi_q
    =
    \frac{1}{L^2 \beta} \sumprime_k \\
    \tr{
        \mathcal{G}_{k+q}
        \mathbb{V}^{(\Phi)}_{k+q,k}
        \mathcal{G}_{k}
        \mathbb{V}^{(\bAy)}_{k,k+q}
    },
\end{equation}
i.e.,
\begin{equation}
    \Ay_{j,-q} (\Pi^{j}_q)^T \Phi_q
    =
    -e \Ay_{j,-q} \frac{1}{L^2 \beta} \sumprime_k
    \tr{
        \mathcal{G}_{k}
        (v^j_{0, \vb{k}+\frac{1}{2}\vb{q}} \tau^0 + \tilde{v}^j_{\vb{k}+\frac{1}{2}\vb{q}} \tau^z)
        \mathcal{G}_{k+q}
        \mathbb{V}^{(\Phi)}_{k+q, k}
    }.
\end{equation}
At $T \rightarrow 0$, after Matsubara frequency integration one finds Eq.~\eqref{eq:mode_couplings_general} of the main text,
\begin{equation}
    \Pi^{j}_{\vb{q}=0}(i\Omega_m)
    =
    -e \int' \frac{\dd^2\vb{k}}{(2\pi)^2}
    \frac{4 \Deltamf}{E_{0,\vb{k}}}
    \frac{\tilde{v}^j_{\vb{k}}}{(i \Omega_m)^2 - 4 E_{0,\vb{k}}^2}
    \begin{pmatrix}
        \xi_{0,\vb{k}} \left|\chi^{(1)}_{\vb{k}}\right|^2
        \\
        \Re \left[
            \overline{\chi^{(1)}_{\vb{k}}}
            \chi^{(2)}_{\vb{k}}
            ( \xi_{0,\vb{k}} + \frac{1}{2} i \Omega_m )
        \right]
        \\
        - \Im \left[
            \overline{\chi^{(1)}_{\vb{k}}}
            \chi^{(2)}_{\vb{k}}
            ( \xi_{0,\vb{k}} + \frac{1}{2} i \Omega_m )
        \right]
    \end{pmatrix}.
\end{equation}

Assuming $C_3$ symmetry, the Higgs mode must remain dark: trigonal warping yields $\tilde{v}^{j}_{\vb{k}} \sim \cos(2\varphi_{\vb{k}}), \sin(2\varphi_{\vb{k}})$, but $\xi_{0,\vb{k}}$ cannot have a component with the same angular periodicity without breaking $C_3$. More generally, $\tilde{v}^j$ and $\xi_0$ cannot have matching angular harmonics unless $\xi = \xi_0 + \tilde{\xi}$ breaks all crystalline rotational symmetries. On the other hand, even in our simple model, the $a$ and $b$ modes become bright:
\begin{multline}
    \Pi^{j \phi}_{\vb{q}=0}(i\Omega_m)
    =
    -e \int' \frac{\dd{k} k}{2 \pi} \int_0^{2\pi} \frac{\dd{\varphi}}{2 \pi}
    \frac{4 \Deltamf}{E_{0,k}}
    \frac{
        \eta k^2 
        [\cos(2\varphi) \delta_{jx} 
        - \sin(2 \varphi) \delta_{jy}]
    }{(i \Omega_m)^2 - 4 E_{0,k}^2}
    \begin{pmatrix}
        \xi_{0,k}
        \\
        \xi_{0,k} \cos(2\varphi)
        - \frac{1}{2} \Omega_m \sin(2\varphi)
        \\
        - \frac{1}{2} \Omega_m \cos(2\varphi)
        + \xi_{0,k} \sin(2\varphi)
    \end{pmatrix}
    \\ \\ =
    -e \int' \frac{\dd{k} k}{2 \pi}
    \frac{2 \Deltamf}{E_{0,k}}
    \frac{\eta k^2}{(i \Omega_m)^2 - 4 E_{0,k}^2}
    \begin{pmatrix}
        0
        \\
        \xi_{0,k} \delta_{j x}
        - \frac{1}{2} \Omega_m \delta_{j y}
        \\
        - \frac{1}{2} \Omega_m \delta_{j x}
        + \xi_{0,k} \delta_{j y}
    \end{pmatrix}
    \\ \\ \approx
    -4 e \nu \eta m \Deltamf
    \int_{-\Lambda}^{\Lambda} \dd{\xi}
    \frac{1}{\sqrt{\xi^2 + \Deltamf^2}}
    \frac{\xi + \mu}{
        (i \Omega_m)^2 - 4 (\xi^2 + \Deltamf^2)
    }
    \begin{pmatrix}
        0
        \\
        \xi \delta_{j x}
        - \frac{1}{2} \Omega_m \delta_{j y}
        \\
        - \frac{1}{2} \Omega_m \delta_{j x}
        + \xi \delta_{j y}
    \end{pmatrix}.
\end{multline}
Taking a large cutoff $\Lambda \gg \Deltamf$ we obtain the results presented in the main text,
\begin{subequations}
\begin{equation}
    \Pi^{x a}_{\vb{q}=0}(i\Omega_m)
    =
    -\Pi^{y b}_{\vb{q}=0}(i\Omega_m)
    =
    2 e \nu \eta m \Deltamf \left(
        \log \frac{2 \Lambda}{\Deltamf}
        - \frac{\sqrt{4 \Deltamf^2 - (i \Omega_m)^2}}{i \Omega_m}
        \arctan \frac{i \Omega_m}{\sqrt{4 \Deltamf^2 - (i\Omega_m)^2}}
    \right)
\end{equation}
and
\begin{equation}
    \Pi^{x b}_{\vb{q}=0}(i\Omega_m)
    =
    \Pi^{y a}_{\vb{q}=0}(i\Omega_m)
    =
    i 2 e \nu \eta m \mu
    \frac{2 \Deltamf}{\sqrt{4 \Deltamf^2 - (i \Omega_m)^2}} \arctan \frac{i \Omega_m}{\sqrt{4 \Deltamf^2 - (i\Omega_m)^2}}.
\end{equation}
\end{subequations}

The $h$, $a$, and $b$ modes do not contribute to the total charge density, since
\begin{equation}    \label{eq:mode_charges}
    (\Pi^0_q)^T \Phi_q
    =
    \frac{i e}{L^2 \beta} \sumprime_k
    \tr{
        \mathcal{G}_{k}
        \tau^z
        \mathcal{G}_{k+q}
        \mathbb{V}^{(\Phi)}_{k+q, k}
    },
\end{equation}
with
\begin{equation}
    \Pi^0_{\vb{q}=0}(i\Omega_m)
    =
    -i e \int \frac{\dd^2{\vb{k}}}{(2\pi)^2}
    \frac{2 \Deltamf}{E_{0,\vb{k}}} \frac{1}{(i \Omega_m)^2 - 4 E_{0,\vb{k}}^2}
    \begin{pmatrix}
        2 \xi_{0,\vb{k}} |\chi^{(1)}_{\vb{k}}|^2 \\
        \Re[
            \overline{\chi}^{(1)}_{\vb{k}} \chi^{(2)}_{\vb{k}} 
            (2 \xi_{0,\vb{k}} + i \Omega_m) 
        ] \\
        - \Im[
            \overline{\chi}^{(1)}_{\vb{k}} \chi^{(2)}_{\vb{k}} 
            (2 \xi_{0,\vb{k}} + i \Omega_m) 
        ]
    \end{pmatrix}
    =
    0.
\end{equation}
The first ($h$) component vanishes by particle-hole symmetry, and the $a$ and $b$ components vanish under the angular integral.

\subsubsection{c) Integrating out the collective modes}
Integrating out the $a$ and $b$ modes renormalizes the superfluid stiffness as per Eq.~\eqref{eq:response_renormalization} in the main text. In our simple model, their contribution is
\begin{multline}
    K^{ij}_{\text{cm}, \vb{q}=0}(i\Omega_m)
    =
    \frac{1}{4} [\Pi^{i}_{\vb{q}=0}(-i\Omega_m)]^T
    \hat{\mathcal{D}}_{\vb{q}=0}(i\Omega_m)
    \Pi^{j}_{\vb{q}=0}(i\Omega_m)
    \\
    =
    \frac{1}{4} \mathcal{D}_{\vb{q}=0}(i\Omega_m) \left[
        \Pi^{i a}_{\vb{q}=0}(-i\Omega_m)
        \Pi^{j a}_{\vb{q}=0}(i\Omega_m)
        +
        \Pi^{i b}_{\vb{q}=0}(-i\Omega_m)
        \Pi^{j b}_{\vb{q}=0}(i\Omega_m)
    \right]
\end{multline}
where the clapping-mode propagator is $\mathcal{D}(i\Omega_m) \equiv \hat{\mathcal{D}}_{aa; \vb{q}=0}(i\Omega_m) = \hat{\mathcal{D}}_{bb; \vb{q}=0}(i\Omega_m)$, given in Eq.~\eqref{eq:mode_propagator_matrix}.

We now analytically continue to real frequency, $i \Omega_m \rightarrow \Omega + i 0^+$. Denoting
\begin{equation}
    f = \frac{\delta g / g_1}{1 - \delta g / g_1}
\end{equation}
and again using $\sin z = \frac{\Omega + i 0^+}{2 \Deltamf}$, we have the result
\begin{multline}
    K^{ij}_{\text{cm}, \vb{q}=0}(\Omega)
    =
    e^2 \nu \frac{(\eta m \mu)^2}{
        \frac{z}{2} (\tan z - \cot z) 
        - f \log(\frac{2 \Lambda}{\Deltamf})
    }
    \\ \times
    \begin{pmatrix}
        \left(z \sec z \right)^2
        + \left( \frac{\Deltamf}{\mu} \right)^2 \left(
            z \cot z
            - \log \frac{2 \Lambda}{\Deltamf}
        \right)^2
        & -i 2 \frac{\Deltamf}{\mu} z
        \left(
            z \csc z
            -
            \sec z \log \frac{2 \Lambda}{\Deltamf}
        \right)
        \\ i 2 \frac{\Deltamf}{\mu} z
        \left(
            z \csc z
            -
            \sec z \log \frac{2 \Lambda}{\Deltamf}
        \right)
        & \left( z \sec z \right)^2 
        + \left( \frac{\Deltamf}{\mu} \right)^2 \left(
            z \cot z
            - \log \frac{2 \Lambda}{\Deltamf}
        \right)^2
    \end{pmatrix}.
    \label{eq:app_Ktensor}
\end{multline}
Here we abuse notation by including the matrix indices $ij$ on $K^{ij}_{\text{cm}}$ to clarify that we are focused only on the spatial part of the response tensor. Note that we found $\Pi^{0}_{\vb{q}=0} = 0$; the $h$, $a$ and $b$ modes do not contribute to the total charge density, and hence the long-wavelength compressibility, in our model. Recall also that one must also account for the fermionic contribution to the longitudinal components as per Eq.~\eqref{eq:realtime_fermion_stiffness}.

\subsection{8. Integrating out the phase mode}
Since our approach does not account for the gauge field $A$ when determining the mean-field saddle point, obtaining a gauge-invariant response in general requires us to integrate over phase fluctuations~\cite{lutchyn2008gaugeinvariantchiral,2016boyackgauge,boyack2020multiplemodes}, i.e., the ABG mode. For $\vb{q} \ne 0$, the ABG mode contributes to the current; this is the usual $\grad \theta$ supercurrent term. At $\vb{q}=0$ there is of course no such term. Formally, we can see that the ABG mode drops out of the $\vb{q}=0$ response by integrating it out.

Recall then that $\Ay_{\mu}(x) = A(x) + \frac{1}{2e} (\partial_{\tau} \theta, -\grad{\theta})$ so $\Ay_{\mu,q} = A_{\mu, q} - \frac{i}{2e} q_{\mu} \theta_q$. After integrating out the fermion and the $\Phi$ modes, the effective action is
\begin{multline}    \label{eq:phase_and_gauge_action}
    S_{\text{eff}} [A,\theta]
    =
    L^2 \beta  \sum_q
    \Ay_{\mu, -q} K^{\mu \nu}_q \Ay_{\nu, q}
    \\ =
    L^2 \beta  \sum_q \left[
        A_{\mu,-q} K^{\mu \nu}_q A_{\nu,q}
        + \frac{i}{2e} \left(
            \theta_{-q} q_{\mu} 
            K^{\mu \nu}_q A_{\nu,_q}
            -
            A_{\mu,-q} K^{\mu \nu}_q
            q_{\nu} \theta_q
        \right)
        + \frac{1}{4e^2}
        \theta_{-q} q_{\mu} K^{\mu \nu}_q
        q_{\nu} \theta_q
    \right].
\end{multline}
Integrating out $\theta$ then yields a contribution to the response tensor of the form
\begin{equation}
    K^{\mu \nu}_{\theta,q}
    =
    - \frac{
        K^{\mu \alpha} q_{\alpha} 
        q_{\beta} K^{\beta \nu}
    }{
        q_{\gamma} K^{\gamma \delta} q_{\delta}
    }
    \xrightarrow{\vb{q}\rightarrow0}
    - \frac{K^{\mu 0} K^{0 \nu}}{K^{00}}.
\end{equation}
Any rotational symmetry $C_{n\ge2}$ will enforce $K^{0 j}_{\vb{q}=0} = K^{j 0}_{\vb{q}=0} = 0$, so $K^{ij}_{\theta,\vb{q}=0} = 0$: the ABG mode drops out of the optical response at $\vb{q}=0$.

\subsection{9. Limits of interest}
Here we obtain approximate expressions for the superfluid response in the DC ($\Omega \rightarrow 0$), collective-mode resonant ($\Omega \rightarrow \Omega_{\text{c}}$), and gap-edge resonant ($\Omega \rightarrow 2 \Deltamf$) limits.

At DC:
\begin{equation}
    K^{i j}_{\text{f}, \vb{q}=0} (\Omega \rightarrow 0)
    =
    \delta_{ij} e^2 \nu
    \left\lbrace
        \frac{\mu}{m}
        - 2 (\eta m \mu)^2
        \left[
             1
            - \left( \frac{\Deltamf}{\mu} \right)^2 
            \left( 
                1
                + \log \frac{2 \Lambda}{\Deltamf}
            \right)
        \right]
    \right\rbrace
\end{equation}
and
\begin{equation}
    K^{ij}_{\text{cm}, \vb{q}=0}(\Omega \rightarrow 0)
    =
    2 \times \delta_{ij} e^2 \nu (\eta m \Deltamf)^2
    \frac{\left(\log \frac{2 \Lambda}{\Deltamf} - 1 \right)^2}{
        1 + 2 f \log \frac{2 \Lambda}{\Deltamf}
    }.
\end{equation}

The collective-mode propagator diverges at the clapping mode resonance, $\mathcal{D}^{-1} (\Omega_c) = 0$, while all other quantities remain finite. Near resonance, we expand the denominator as
\begin{equation}
    \frac{z}{2} (\tan z - \cot z) - f \log \frac{2 \Lambda}{\Deltamf}
    =
    (\Omega - \Omega_{\text{c}} + i 0^+) \frac{\partial}{\partial \Omega} \left[
        \frac{z}{2} (\tan z - \cot z)
    \right]_{\Omega = \Omega_{\text{c}}}
    + O \left[(\Omega - \Omega_{\text{c}})^2 \right],
\end{equation}
i.e.,
\begin{equation}
    \frac{z}{2} (\tan z - \cot z) - f \log \frac{2 \Lambda}{\Deltamf}
    \approx
    -\frac{\Omega - \Omega_{\text{c}} + i 0^+}{\Omega_{\text{c}}} \left[
        1 - \frac{1}{2} \sec^2 z_{\text{c}} \left(
            1 + z_{\text{c}} \sec z_{\text{c}} \csc z_{\text{c}}
        \right)
    \right]
\end{equation}
where $z_{\text{c}} \equiv z(\Omega = \Omega_{\text{c}})$. Substituting $\Omega \rightarrow \Omega_{\text{c}}$ everywhere else yields
\begin{multline}
    K^{ij}_{\text{cm}, \vb{q}=0}(\Omega \approx \Omega_{\text{c}})
    =
    -e^2 \nu \frac{(\eta m \mu)^2 \Omega_{\text{c}}}{
        \Omega - \Omega_{\text{c}} + i 0^+
    } \frac{1}{
        1 - \frac{1}{2} \sec^2 z_{\text{c}} (
            1 + z_{\text{c}} \sec z_{\text{c}} \csc z_{\text{c}}
        )
    }
    \\ \times
    \begin{pmatrix}
        \left(z_{\text{c}} \sec z_{\text{c}} \right)^2
        + \left( \frac{\Deltamf}{\mu} \right)^2 \left(
            z_{\text{c}} \cot z_{\text{c}}
            - \log \frac{2 \Lambda}{\Deltamf}
        \right)^2
        & -i 2 \frac{\Deltamf}{\mu} z_{\text{c}}
        \left(
            z_{\text{c}} \csc z_{\text{c}}
            -
            \sec z_{\text{c}} \log \frac{2 \Lambda}{\Deltamf}
        \right)
        \\ i 2 \frac{\Deltamf}{\mu} z_{\text{c}}
        \left(
            z_{\text{c}} \csc z_{\text{c}}
            -
            \sec z_{\text{c}} \log \frac{2 \Lambda}{\Deltamf}
        \right)
        & \left( z_{\text{c}} \sec z_{\text{c}} \right)^2 
        + \left( \frac{\Deltamf}{\mu} \right)^2 \left(
            z_{\text{c}} \cot z_{\text{c}}
            - \log \frac{2 \Lambda}{\Deltamf}
        \right)^2
    \end{pmatrix}.
\end{multline}
The result simplifies drastically when $f = 0$, i.e., $g_1 = g_2$, in which case $\Omega_{\text{c}} = \sqrt{2} \Deltamf$, so $z_{\text{c}} = \arcsin{\frac{1}{\sqrt{2}}} = \pi/4$, $\sin z_{\text{c}} = 1/\csc z_{\text{c}} = \cos z_{\text{c}} = 1 / \sec z_{\text{c}} = 1 / \sqrt{2}$, and $\tan z_{\text{c}} = \cot z_{\text{c}} = 1$. The divergent (collective-mode) part of the superfluid stiffness tensor is then
\begin{equation}
    K^{ij}_{\text{cm}, \vb{q}=0}(\Omega \approx \sqrt{2} \Deltamf)
    \approx
    \frac{2 e^2 \nu}{\pi} (\eta m \mu)^2 \frac{ \sqrt{2}\Deltamf}{
        \Omega - \sqrt{2} \Deltamf + i 0^+
    }
    \begin{pmatrix}
        \frac{\pi^2}{8}
        + \left(
            \frac{\Deltamf}{\mu}
            \log \frac{2 \Lambda}{\Deltamf}
        \right)^2
        & i \frac{\Deltamf}{\mu} \frac{\pi}{\sqrt{2}}
        \log \frac{2 \Lambda}{\Deltamf}
        \\ -i \frac{\Deltamf}{\mu} 
        \frac{\pi}{\sqrt{2}}
        \log \frac{2 \Lambda}{\Deltamf}
        & \frac{\pi^2}{8}
        + \left(
            \frac{\Deltamf}{\mu}
            \log \frac{2 \Lambda}{\Deltamf}
        \right)^2
    \end{pmatrix}.
\end{equation}

Lastly we have the gap-edge feature, which involves a subtle cancellation between the bare bubble and collective-mode contributions; both diverge separately, but their divergences cancel one another. One finds
\begin{multline}
    K^{xx}_{\vb{q}=0} (\Omega \approx 2 \Deltamf)
    =
    e^2 \nu \Bigg\lbrace
        \frac{\mu}{m}
        + 2 (m \eta \mu)^2 \left(2 f - \frac{\Deltamf^2}{\mu^2} \right) \log \frac{2 \Lambda}{\Deltamf}
        \\ + (m \eta \mu)^2 \left[
            \pi \left( 1 + \frac{\Deltamf^2}{\mu^2} \right)
            + \frac{4}{\pi} \left(
                4 f^2 + \frac{\Deltamf^2}{\mu^2} 
            \right) \left( \log \frac{2 \Lambda}{\Deltamf} \right)^2
        \right] \sqrt{\frac{2 \Deltamf - \Omega - i 0^+}{\Deltamf}}
    \Bigg\rbrace
    + O\left( \frac{2 \Deltamf - \Omega}{\Deltamf} \right)
\end{multline}
and
\begin{multline}
    K^{xy}_{\vb{q}=0} (\Omega \approx 2 \Deltamf)
    =
    2 i e^2 \nu (\eta m \mu)^2 \frac{\Deltamf}{\mu} \left\lbrace
        2 \log \frac{2 \Lambda}{\Deltamf}
        + \left[ \frac{8}{\pi} f \left( \log \frac{2 \Lambda}{\Deltamf} \right)^2 - \pi \right]
        \sqrt{\frac{2 \Deltamf - \Omega - i 0^+}{\Deltamf}}
    \right\rbrace
    \\ + O \left( \frac{2 \Deltamf - \Omega}{\Deltamf}\right).
\end{multline}
Recall that above the gap edge ($\Omega > 2 \Deltamf$), $\sqrt{2 \Deltamf - \Omega - i 0^+} = -i \sqrt{\Omega - 2 \Deltamf}$. Just below the gap, the longitudinal response is purely real, and the transverse response purely imaginary. The longitudinal (transverse) conductivity is then purely imaginary (real); recall that $\sigma = i K / (\Omega + i 0^+)$. Crossing the gap edge from below, the longitudinal (transverse) response develops an imaginary (real) part with a square-root onset, so the longitudinal (transverse) conductivity develops a real (imaginary) part.

\subsection{10. Inductance}
Our results demonstrate that inductance measurements could detect the clapping mode in trigonally-warped single-valley superconductors. We can see this by expressing the action in terms of the supercurrent. Starting from
\begin{equation}
    \frac{1}{L_x L_y \beta} S_{\text{eff}}
    =
    \sum_q \Ay_{i,-q} K^{ij}_q \Ay_{j,q}
\end{equation}
(recall $\bAy = \vb{A} - \frac{\hbar}{2e} \grad \theta$), the supercurrent density is
\begin{equation}
    J^i_q = - \frac{1}{L_x L_y \beta} \frac{\delta S_{\text{eff}}}{\delta A_{i,-q}}
    =
    \left(
        K^{ij}_q + K^{ji}_{-q}
    \right) \Ay_{j,q}
    =
    2 K^{ij}_q \Ay_{j,q}
    \qquad \text{so} \qquad
    \bAy_q = \frac{1}{2} K^{-1}_q \vb{J}_q
\end{equation}
where we used $K_q^{ij} = K_{-q}^{ji}$. Throughout this subsection, we restrict $K$ to its spatial components (ignoring $K^{0\mu}$ and $K^{\mu 0}$). In 2D, the current density is per unit transverse direction. Writing $J^{x/y} = \frac{I^{x/y}}{L_{y/x}}$, i.e.,
\begin{equation}
    \vb{J} = \begin{pmatrix}
        1/L_y & 0 \\
        0 & 1/L_x
    \end{pmatrix} \vb{I},
\end{equation}
we have
\begin{equation}
    \frac{1}{\beta} S_{\text{eff}}
    =
    \frac{1}{4} L_x L_y \sum_q 
    \vb{J}_{-q}^T K^{-1}_q \vb{J}_q
    =
    \frac{1}{4} \sum_q
    \vb{I}_{-q} \begin{pmatrix}
        L_x & 0 \\
        0 & L_y
    \end{pmatrix} K^{-1}_q
    \begin{pmatrix}
        1/L_y & 0 \\
        0 & 1/L_x
    \end{pmatrix} \vb{I}_q.
\end{equation}
We recognize the free energy of an inductor $U = \frac{1}{2} \mathfrak{L} I^2$, with frequency-dependent inductance matrix
\begin{equation}
    \mathfrak{L} (\Omega)
    =
    \frac{1}{2} \begin{pmatrix}
        L_x & 0 \\
        0 & L_y
    \end{pmatrix} K^{-1}(\Omega)
    \begin{pmatrix}
        1/L_y & 0 \\
        0 & 1/L_x
    \end{pmatrix}.
\end{equation}
Assuming a square sample ($L_x = L_y$), we have simply $\mathfrak{L}(\Omega) = \frac{1}{2} K^{-1}(\Omega)$.

\subsection{11. Numerical estimates}

In this Section we provide order-of-magnitude estimates for the model parameters pertinent to R$n$G. We caution that a quantitative treatment taking into account the effect of electronic interactions, in addition to details of the band structure, will be necessary to interpret microwave experiments.

Comparing to band structure and Hartree-Fock calculations in the literature, we estimate parameters relevant to the electron-doped region of the phase diagram at relatively high displacement field, where chiral superconductivity is reported in rhombohedral tetralayer and pentalayer graphene. We take $\mu \sim 1$ meV~\cite{yoon2025quartermetalsuperconductivity}, and a conservative estimate for the density of states (per spin and valley flavor) $\nu \sim 5 \text{~eV}^{-1} \text{nm}^{-2}$ close to van Hove singularities~\cite{Ghazaryan2023}. The effective trigonal warping term is more delicate as it sensitively depends on the geometry and topology (simply-connected, annular or disconnected) of the Fermi surface, which can in turn be modified by electronic interactions and symmetry breaking.

We provide a rough estimate of the effective trigonal warping term by projecting the non-interacting, continuum-model description of rhombohedral $L$-layer graphene (which contains hopping terms $\gamma_{0,1,2,3}$ useful below) to the two sublattices ($A_1$,$B_L$) where the low-energy wavefunctions are predominantly polarized. Following the strategy outlined in Ref.~\cite{Slizovskiy2019}, the dispersion of the lowest valence band reads
\begin{equation}
    \epsilon(\bk) = \frac{\hbar^2 \bk^2}{2m} + \sqrt{ u^2 + \gamma_1^2 X^\dagger(\bk) X(\bk)} ,
    \label{eq:dispersion_projectedmodel}
\end{equation}
where $u$ denotes (half of) the electrostatic potential between the ($A_1$,$B_L$) sublattices, and the function $X$ combines various contributions up to $L$th order in $\bk$. For $L=4$ and considering the $K$ valley for concreteness, the lowest-order terms of interest read
\begin{equation}
    X(\bk) = \frac{\gamma_2}{\gamma_1} \frac{k_x - i k_y}{k_0} + \frac{\gamma_3^2}{\gamma_0^2} \frac{(k_x + i k_y)^2}{k_0^2} + \mathcal{O}(\vb{k}^3, \vb{k}^4),
\end{equation}
with the parameter $k_0 = 2 \gamma_1 / \sqrt{3} a_0 \gamma_0$.
The product $X^\dagger(\bk) X(\bk)$ yields contributions up to 8th-order in momentum. The second-order contribution will simply renormalize the effective mass $m$ and we therefore neglect it. Focusing on the cubic-in-momentum contribution, relevant for trigonal warping, one finds
\begin{equation}
    X^\dagger(\bk) X(\bk) = \frac{2 \gamma_2 \gamma_3^2}{\gamma_1 \gamma_0^2} \frac{\left(k_x^3 - 3 k_x k_y^2 \right)}{k_0^3} + \mathcal{O}(\bk^4).
\end{equation}
In the limit of large displacement field $u$, we can expand Eq.~\ref{eq:dispersion_projectedmodel} and obtain the dispersion relation (up to 3th order in $\vb{k}$)
\begin{equation}
    \epsilon(\bk) = u + \frac{\hbar^2 \bk^2}{2m} +  \frac{\gamma_1 \gamma_2 \gamma_3^2}{\gamma_0^2 u} \frac{\left(k_x^3 - 3 k_x k_y^2 \right)}{k_0^3} .
\end{equation}
Using numerical values $\gamma_0 = 3.1$eV, $\gamma_1 = 380$meV, $\gamma_2 = -15 $meV and $\gamma_3 = -290$ meV commonly used in the literature to model multilayer graphene~\cite{Ghazaryan2023} and an interlayer potential $u=40$ meV, we have $k_0 \approx 0.58$ nm$^{-1}$. Comparing to the dispersion relation in Eq.~\ref{eq:app_dispersion_trigonal}, we extract the trigonal warping parameter
\begin{equation}
    \eta = \frac{3 \gamma_1 \gamma_2 \gamma_3^2}{\gamma_0^2 u k_0^3} \approx  - 20 \text{~meV~nm}^3 .
\end{equation}
\end{widetext}

\end{document}